\documentclass[%
reprint,
superscriptaddress,
 amsmath,amssymb,
 aps,
 prf,
]{revtex4}
\usepackage{graphicx}
\usepackage{dcolumn}
\usepackage{bm}
\usepackage{color}
\usepackage{soul}
\usepackage{array}
\usepackage{nomencl}
\usepackage{float}
\usepackage{multirow}
\usepackage{algorithm,algpseudocode}

\usepackage{xcolor}
\usepackage{ulem}
\graphicspath{{figures/}}

\newcommand{\ml}[1]{{{\color{black}#1}}}

\definecolor{darkgreen}{rgb}{0.0, 0.5, 0.0}

\usepackage{subcaption}
\begin{document}

\preprint{APS/123-QED}

\title{Regime identification for stratified wakes from limited measurements: \ml{a library-based sparse regression formulation}}

\author{Vamsi Krishna Chinta}
\email{vchinta@usc.edu}
\affiliation{Aerospace and Mechanical Engineering,\\
University of Southern California, Los Angeles, CA 90089}

\author{Chan-Ye Ohh}
\affiliation{Aerospace and Mechanical Engineering,\\
University of Southern California, Los Angeles, CA 90089}

\author{Geoffrey Spedding}
\affiliation{Aerospace and Mechanical Engineering,\\
University of Southern California, Los Angeles, CA 90089}

\author{Mitul Luhar}
\affiliation{Aerospace and Mechanical Engineering,\\
University of Southern California, Los Angeles, CA 90089}



\begin{abstract}
Bluff body wakes in stratified fluids are known to exhibit a rich range of dynamic behavior that can be categorized into different regimes based on Reynolds number ($Re$) and Froude number ($Fr$). 
Topological differences in wake structure across these different regimes have been clarified recently through the use of Dynamic Mode Decomposition (DMD) on Direct Numerical Simulation (DNS) and laboratory data for a sphere in a stratified fluid for $Re\in [200,1000]$ and $Fr\in[0.5,16]$. In this work, we attempt to identify the dynamic regime from limited measurement data in a stratified wake with (nominally) unknown $Re$ and $Fr$. A large database of candidate basis functions is compiled by pooling the DMD modes obtained in prior DNS. A sparse model is built using the Forward Regression with Orthogonal Least Squares (FROLS) algorithm, which sequentially identifies DMD modes that best represent the data and calibrates their amplitude and phase. After calibration, the velocity field can be reconstructed using a weighted combination of the dominant DMD modes. The dynamic regime for the measurements is estimated via a projection-weighted average of $Re$ and $Fr$ corresponding to the identified modes. Regime identification is carried out from a limited number of 2D velocity snapshots from numerical and experimental datasets, as well as 3 point measurements in the wake of the body. A metric to assess confidence is introduced based on the observed predictive capability. This approach holds promise for the implementation of data-driven fluid pattern classifiers.
\end{abstract}

\maketitle


\section{Introduction}\label{sec:intro}
\subsection{Motivation and problem statement}
Wakes behind bluff bodies in natural environments frequently evolve in the presence of stable stratification in fluid density. Such stratified wakes can be long lived and coherent \citep{spedding2014wake}. The topology of these wakes can contain valuable information regarding the upstream bluff body as well as the flow regime. Indeed, many aquatic organisms are thought to use the cues from wake signatures for navigation. For instance, it has been proposed that Brazilian green sea turtles make use of wake signatures from islands to reliably make a round trip of approximately 4400 km \citep{papi1996pinpointing}. Fish, seals, and copepods also use such navigational cues for mate finding, predator evasion, and hunting \citep{montgomery1997lateral,ristroph2015lateral}. In general, the ability to identify the wake creator from limited measurements in the wake could aid object identification, navigation, and motion planning efforts in environments with, and without stratification.

Stratified wakes are influenced by inertial, viscous, and buoyancy forces. Hence, the relevant non-dimensional parameters governing these flows are the Reynolds number, $Re$, which is the ratio between the inertial and viscous forces, and the Froude number, $Fr$, which is the ratio between the convective time scale and the buoyancy time scale. Perturbations to a stably stratified fluid give rise to internal waves of characteristic frequency $N$, which is called the Brunt-V{\"a}is{\"a}l{\"a} frequency. This characteristic frequency is defined as $N = \sqrt{-\frac{g}{\rho_{o}} \frac{\partial\overline{\rho}}{\partial z}}$, where $g$ is the acceleration due to gravity, $\overline{\rho}$ is the fluid density which decreases in the vertical ($z$) direction, and $\rho_{o}$ is a mean density. Compared to non-stratified wakes, buoyancy effects in stratified wakes give rise to a rich variety of topological structures \citep{lin1992stratified,chomaz1993structure}. 
Stable density stratification suppresses turbulent fluctuations in the vertical direction \citep{lin:79, spedding:01, brucker:10, redford:15}, while simultaneously enhancing the possibility of vertical momentum and energy transport through the internal waves excited either by the body itself, or by the wake. It has also been argued that late wakes from streamlined and sharp-edged bodies evolve in a self-similar manner when scaled appropriately, which suggests that information regarding the initial conditions is lost \citep{meunier2004loss}. However, it is possible this loss of information is limited to statistical quantities and that the wake topology may yet retain information regarding its origins \citep{spedding2014wake}.

\begin{figure}
    \centering
    \includegraphics[width = 0.525\linewidth]{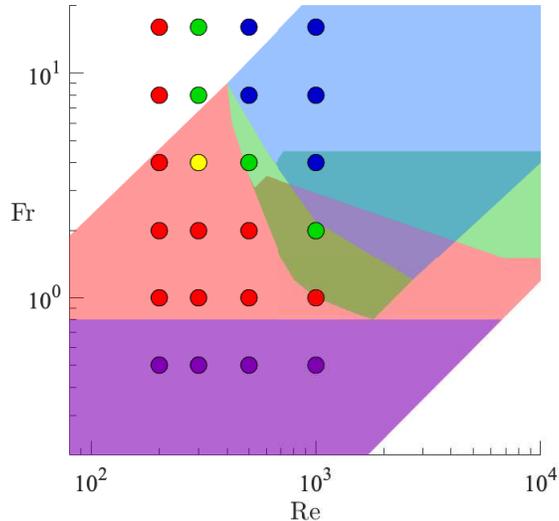}    
    \caption{Regime space in terms of Reynolds number ($Re$) and Froude number ($Fr$). Filled circles show the cases used in this study. \ml{Physical regimes identified in previous studies are denoted by the color of the circles \citep{madison:21,ohh2021wake} and the background shading \citep{lin1992stratified,chomaz1993structure}. Regimes are color-coded as: vortex street (purple), symmetric non-oscillation (red), asymmetric non-oscillation (yellow), planar oscillation (green), and spiral mode (blue).  Overlapping shaded regions reflect differences in regime boundaries identified in previous studies.}}
    \label{fig:ReFrspace}
\end{figure}

\begin{figure}
    \centering
    \includegraphics[width = 0.7\linewidth]{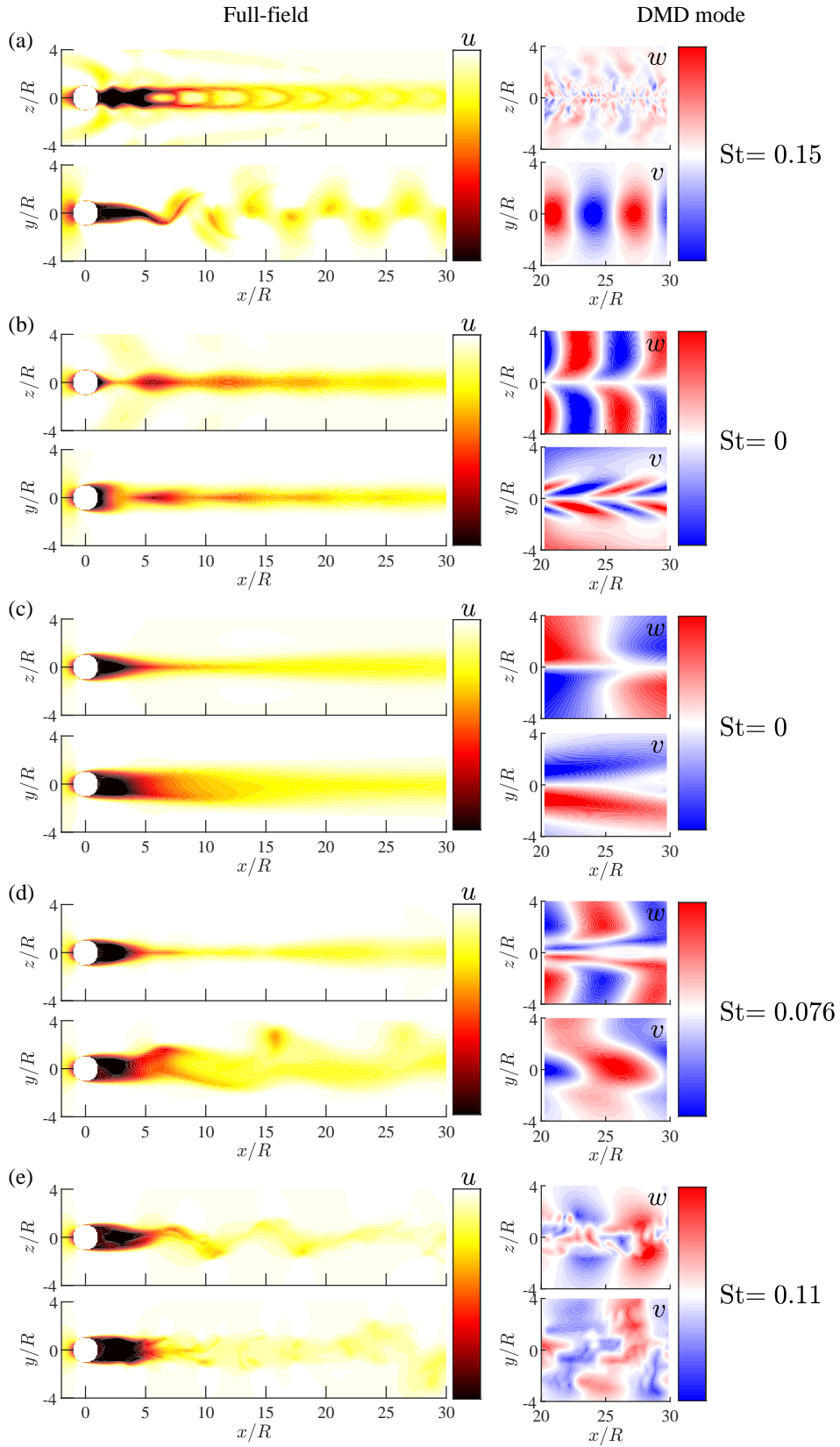}
    \caption{\ml{Snapshots of normalized streamwise velocity ($u$) from DNS (left column) and cross-stream velocity components ($v$, $w$) for the dominant DMD mode (right column) in different regimes: (a) vortex street at $(Re,Fr)=(1000,0.5)$ , (b) symmetric non-oscillation at $(Re,Fr)=(500,1)$, (c) asymmetric non-oscillation at $(Re,Fr)=(300,4)$, (d) planar oscillation at $(Re,Fr)=(500,4)$, and (e) spiral mode at $(Re,Fr)=(1000,8)$. \ml{Shading for the DMD modes shows positive (red) and negative (blue) values for $v$ and $w$, normalized by $0.75 \mathrm{max}(|v|,|w|)$. The Strouhal number, $St = f D/ U_{o}$, for the oscillation frequency $f$ of the DMD modes is provided.}}}
    \label{fig:velContours_regimes}
\end{figure}

\ml{Several previous experimental and numerical studies have characterized the topology of stratified wakes as a function of $Re$ and $Fr$ \citep{hanazaki1988numerical,lin1992stratified,chomaz1993structure,orr2015numerical,pal2016regeneration,chongsiripinyo2017vortex,pal2017direct}. Recent work by \citet{ohh2021wake} and \citet{madison:21} proposes that stratified sphere wakes can be classified into five principal regimes across $(Re,Fr)$ space.  The Reynolds number is $Re = U_o D/\nu$ and the Froude number is defined (based on radius) as $Fr = U_o/(NR)$. Here, $U_{o}$ is the freestream velocity, $D$ is sphere diameter, $R$ is the radius, and $\nu$ is the kinematic viscosity of the fluid. As shown in figure~\ref{fig:ReFrspace}, these five regimes are labelled: vortex street, symmetric non-oscillation, asymmetric non-oscillation, planar oscillation, and spiral mode, describing the qualitatively different modes that emerge from the near-wake.  Figure~\ref{fig:ReFrspace} includes the regime boundaries identified by \citet{lin1992stratified} and \citet{chomaz1993structure} based on the body and near-wake flow features. Though the regimes delineated in these studies involved a dense coverage of parameter space that implied sharp transitions between regimes, the boundaries and regimes themselves are not identical and we remain agnostic on whether or not transitions are sharp.  Here the $(Re,Fr)$ space is populated by experimental and numerical data at 24 discrete combinations reported in \cite{ohh2021wake}, and shown using filled circles in figure~\ref{fig:ReFrspace}.} 
\ml{Topological distinctions between the different regimes are illustrated in figure~\ref{fig:velContours_regimes}.

For each regime, figure~\ref{fig:velContours_regimes} shows contours of streamwise velocity in the horizontal ($x-y$) and vertical ($x-z$) planes obtained in DNS, along with cross-stream velocity components for the dominant DMD mode (see Sec.~\ref{sec:methods} for further detail on the simulations and DMD). The vortex street regime is illustrated in figure~\ref{fig:velContours_regimes}(a).  Buoyancy forces due to stratification are the strongest in this regime ($Fr=0.5$ is the lowest value) and excursions in the vertical plane are strongly suppressed.  The dominant DMD mode in figure~\ref{fig:velContours_regimes}(a) shows the presence of a coherent vortex street in the horizontal plane while the slice in the vertical plane shows low amplitude traces of internal waves. When $Fr \approx 1$, the lee wave is steady in the body reference frame and dominates the near wake. Fluctuations are suppressed in both the horizontal and vertical planes, which results in a time-invariant flow field that is classified as symmetric non-oscillation (figure~\ref{fig:velContours_regimes}(b)). With further increases in $Fr$, an asymmetry appears about the vertical plane for $Re = 300$. This regime, termed asymmetric non-oscillation, is shown in figure~\ref{fig:velContours_regimes}(c).  Though only one instance of this regime is observed in the dataset being used here, at $(Re,Fr)=(300,4)$ it is a consistent feature in both laboratory and numerical experiment. As $Fr$ increases further, the wake contains a mix of horizontal oscillations and vertical motions associated with internal waves  (figure~\ref{fig:velContours_regimes}(d)). This regime is termed planar oscillation because most kinetic energy is associated with the horizontal oscillatory motion.  These oscillations eventually appear in the vertical plane, replacing the internal wave modes, as $Re$ and $Fr$ rise, and the wake becomes fully three dimensional.  This regime is termed the spiral mode (figure~\ref{fig:velContours_regimes}(e)) after the known shedding mode that appears at all higher $Re$ \citet{chomaz1993structure}.}

\ml{The classification system described above is based on known and observed physical mechanisms, but it relies on topological distinctions identified and delineated by humans, who design the classification system}. Here, we develop an \ml{expert}-free and purely data-driven approach to identifying the dynamical regime (as gauged by an estimated $Re$ and $Fr$ pair) from sparse measurements in a stratified wake. This approach assumes the availability of significant prior measurement or simulation data across the relevant $(Re,Fr)$ space. As detailed below, regime identification then requires solution of a \textit{sparse} regression problem that seeks to describe the wake structures present in the input (from an unknown regime) in terms of the wake structures present in the large database of prior measurements or simulations (from known regimes). This approach also enables spatio-temporal \textit{reconstruction} of the wake from limited data.

\subsection{Previous \ml{data-driven} classification and reconstruction efforts}
There have been many prior classification and reconstruction efforts for unstratified wakes behind bluff bodies. For instance, such wakes are often classified into different categories based on the number of alternating vortices shed per oscillation period \citep{williamson1988vortex}. For example, 2S refers to 2 single vortices shed per period of oscillation, 2P refers to 2 pairs of vortices of opposite signs shed period, S+P refers to an asymmetric version where a single vortex and a pair of vortices are shed per period, and so on. Similar to the problem being tackled in the present study, \citet{colvert2018classifying} classified the wake behind a flapping airfoil into different regimes (2S, 2P+2S, etc.) from point measurements of vorticity using neural networks. In a continuing study, \citet{alsalman2018training} evaluated four different kinds of flow sensors to classify the flow using neural networks and concluded that the use of a sensor that measures transverse velocity led to more robust classification. While pre-trained neural networks perform admirably for classification purposes, they have some important limitations. Specifically, their utility is limited to the type of input data used for training purposes.  For instance, a network trained using velocity snapshots cannot be used for point measurements and a network trained using point measurement data may not be used for velocity snapshots.  In addition, such networks have limited ability to reconstruct velocity fields or generate predictions for input signals from regimes not included in the training data. The framework developed in the present study begins to address some of these limitations.

For completeness we note that there have been many prior studies on \ml{super-resolution reconstruction, flow field estimation from wall measurements, and state estimation} from limited \ml{and non-intrusive measurements.} Such studies have employed physics-based models \citep[e.g.][]{gomez2016estimation,beneddine2017unsteady,illingworth2018estimating,krishna2020reconstructing,towne2020resolvent,wang2021model,wang2021state}, data-driven modal representations \citep[e.g.][]{bui2004aerodynamic,willcox2006unsteady,tu2013dynamic,tu2013integration}, as well as machine learning tools \citep[e.g.][]{fukami2019super,erichson2020shallow,fukami2021machine,alsalman2018training,guastoni2021convolutional,guemes2021coarse}. A detailed review of these efforts is not provided here for brevity. \ml{Recently, \citet{graff2020reduced} used an approach similar to the present study to build dynamic models for Poiseuille flow using simulation data, and for a rotating fin using PIV data. Specifically, \citet{graff2020reduced} extended the Least Angle Regression algorithm to select DMD modes from a library that best fit the input data, and used this algorithm to develop reduced-complexity dynamic models. However, we note that these prior efforts on reconstruction, estimation, and the development of reduced-complexity dynamic models do not typically target regime or parameter identification.}

For stratified wakes, a number of experimental \citep{lin1992stratified, chomaz1993structure} and computational \citep{pal2016regeneration,pal2017direct} studies have made detailed descriptions of body-generated wake structures as a function of $Re$ and $Fr$. In recent years, modal analysis techniques such as proper orthogonal decomposition (POD) and dynamic mode decomposition (DMD) have been used to extract coherent structure\ml{s} and to build reduced order models \citep{diamessis:10, xiang:17}. 
Snapshot POD yields spatially orthogonal modes ordered based on energy, with temporal coefficients that may contain a mix of frequencies \citep{taira2017modal}. To isolate modes of specific frequency, modal analysis methods that yield temporally orthogonal modes such as DMD \citep{chen2012variants,schmid2010dynamic} or spatially \textit{and} temporally orthogonal modes such as Spectral POD \citep{towne2018spectral} can be used. 
Previous efforts have successfully implemented both snapshot POD and DMD to extract coherent structures in stratified wakes \citep{ohh2021wake, nidhan2019dynamic}. \citet{nidhan2019dynamic} used DMD to analyze the wake structures from numerical datasets at two different Reynolds numbers with varying buoyancy effects ($Fr$). More recently, \citet{ohh2021wake} used DMD to classify wake structures at various $Re$ and $Fr$ from both numerical and experimental datasets. A custom algorithm informed by the \ml{physically-motivated classification system described in the previous section} was developed to characterize wake topology. \ml{This algorithm interrogated the velocity field, vorticity field, and oscillation frequency associated with the most energetic DMD mode and made use of an expert-defined decision tree to classify wake topology into one of the five different physical regimes shown in figure~\ref{fig:ReFrspace}}. 
To our knowledge, there are no previous works that directly estimate $Re$ and $Fr$ for stratified wakes from limited measurements.

\subsection{Contribution and outline}
In this study, we attempt regime identification and flow field reconstruction from limited measurements in stratified wakes behind a sphere.  We use a purely data-driven approach in which the input data (i.e., a small number of velocity snapshots or point measurements) are represented in terms of DMD modes obtained from numerical simulations across a wide range of $Re$ and $Fr$.
These DMD modes are pooled into a large database. We then seek a sparse representation of the input signal in terms of a linear combination of modes from this database. The forward regression with orthogonal least squares algorithm \citep[FROLS][]{billings2013nonlinear} identifies the dominant DMD modes that best represent the data. In other words, only a few dominant DMD modes are selected to represent the input data. The parameter regime for the input data is then estimated using a weighted average of $Re$ and $Fr$ values for the DMD modes selected. We also evaluate the predictive capability of the flow field model built using this process and introduce a metric to evaluate confidence in the predictions.

There are a few important distinctions between the present study and previous efforts on flow reconstruction and classification.  Flow reconstruction efforts typically use a library of POD or DMD modes from the same parameter regime as the measurements to reconstruct the velocity field in space or time.  Here, we use a library of DMD modes from various known regimes to reconstruct the velocity field using limited input data from an unknown regime. In so doing, we can generate quantitative estimates for $Re$ and $Fr$.  Unlike previous classification efforts, this approach also allows for interpolation in parameter values rather than binning into discrete regimes. We also note that, though the present effort focuses on stratified wakes, the regime identification framework can be generalized to other flows. 

The remainder of the paper is organized as follows. In Sec.~\ref{sec:methods}, we describe the simulations and experiments that generated the data used in this study together with the FROLS algorithm used for sparse regression. We also discuss the approach used to identify the parameter regime, and the metrics used to evaluate prediction accuracy and confidence. Note that we attempt regime identification using a variety of input data from DNS and laboratory experiments. This includes planar velocity snapshots (c.f., 2D measurements from Particle Image Velocimetry, PIV) as well as point measurements.
In Sec.~\ref{sec:results}, we first present the results for the baseline case in which planar velocity fields from DNS are used as the input data and regime identification is carried out using a comprehensive DMD mode database. We then use a sub-sampled database of DMD modes (i.e., a database that has only half the available $Re$ and $Fr$ cases) to evaluate predictive capability and test whether the regime identification framework is capable of interpolation. Next, we attempt regime identification with real-world experimental data as the input.  We first use snapshots of the velocity field obtained from PIV for regime identification, and we evaluate predictive capability using different components of velocity. Next, we consider regime identification from point measurements also attempt to identify sensor location. In other words, the framework is used to not only identify $Re$ and $Fr$ based on the DMD modes selected, but also the physical placement of the sensors on the spatial DMD modes. We also evaluate the effect of database size for input data from 2D2C snapshots and point measurements. Finally, in Sec.~\ref{sec:conclusions} we present a brief summary of our findings and conclu ding remarks.

\section{Methods}\label{sec:methods}
In this section, we first describe the simulations used to generate the DMD database and evaluate the regime identification algorithm. We also describe the laboratory experiments used to generate additional datasets used for classification purposes.  We subsequently present the problem formulation and the FROLS algorithm used to find a sparse representation for the input data. Note that this effort makes use of the same experimental and numerical dataset as \citet{ohh2021wake}, who developed a physics-informed algorithm to classify stratified wakes into the different physical regimes shown in figure~\ref{fig:ReFrspace}.

\subsection{Numerical Simulations and DMD database}\label{sec:numerics}

\begin{figure}
 \centering
 \includegraphics[width=0.95\linewidth]{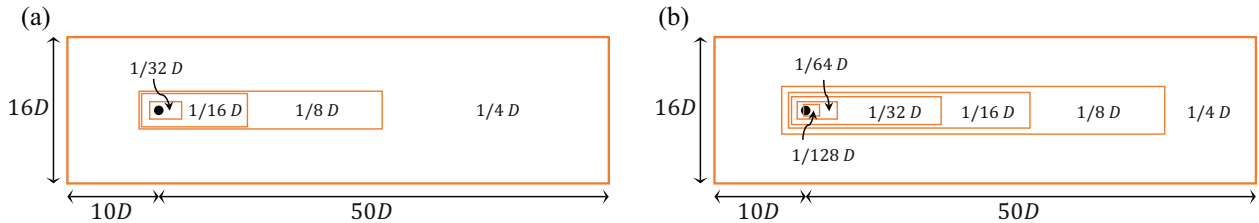}
 \caption{\label{fig:NumMesh} Meshes used for numerical simulations (a) at $Re = \{200,300,500\}$ and (b) at $Re = 1000$.  The mesh used for the lower $Re$ cases has four refinement levels ($2D^{-[2 \cdots 5]}D$). The mesh used for the higher $Re$ case has six refinement levels ($2D^{-[2 \cdots 7]}$).}
 \end{figure}

Direct numerical simulations of stratified wakes behind spheres were carried out systematically over a wide span of the $(Re,Fr)$ parameter space.  Four different Reynolds numbers, $Re = U_{o} D/\nu = \{200, 300, 500, 1000\}$, and six different Froude numbers, $Fr = U_{o} /(NR) = \{0.5, 1, 2, 4, 8, 16\}$, were considered for a total of twenty four different cases (see \ml{filled circles} in figure~\ref{fig:ReFrspace}).  Following \citet{ohh2021wake} and \citet{madison:21}, stratified sphere wake flows in this parameter space can be classified into 5 broad categories, as denoted by the shaded regions in figure~\ref{fig:ReFrspace}. 

The numerical simulations made use of the Boussinesq approximation to the Navier-Stokes equation together with a density transport equation. The dimensionless form of the governing equations is
\begin{equation} \label{NSE-strat}
\nabla \cdot \mathbf{u} = 0,
\end{equation}
\begin{equation}
\frac{\partial{\mathbf{u}}}{\partial{t}} + \mathbf{u} \cdot \nabla \mathbf{u} = -\nabla p' + \frac{1}{Re} \nabla^2 \mathbf{u} + \frac{1}{Fr^2} \rho' \mathbf{g},
\end{equation}
\begin{equation}
\frac{\partial{\rho'}}{\partial{t}} + \mathbf{u} \cdot \nabla \rho' = w \frac{d \overline{\rho}}{dz} + \frac{1}{Pr Re} \nabla^2 \rho',
\end{equation}
where $\mathbf{u}$ is the velocity field, $t$ is time, $\mathbf{g}=-g \mathbf{e}_z$ is the gravitational acceleration in the vertical ($z$) direction, $w$ is the vertical velocity, and $Pr$ is the Prandtl number. The mean background density field is denoted $\overline{\rho}$ and the fluctuating density field is $\rho'$.  Similarly, $p'$ represents the pressure fluctuations about the background hydrostatic pressure.
The governing equations were solved using the PimpleFOAM solver in the OpenFOAM suite\ml{, which uses a finite volume method with a pressure-velocity coupling algorithm. Second-order accurate temporal and spatial discretization schemes were used}. To reduce internal wave reflection from the wall, the $y$-normal and $z$-normal walls of the domain were set to zero-gradient conditions. To allow for possible breaks in the axisymmetry of the wake, the sphere was oscillated once along each of the three axis as an initial condition. The extent of the computational domain was $[-10D, 50D]$ in the streamwise ($x$) direction, $[-8D, 8D]$ in the lateral ($y$) direction, and  $[-8D, 8D]$ in the vertical direction. The position $x=0$, $y=0$, $z=0$ corresponds to the center of the sphere. Multiple grid refinement levels were used \ml{in an unstructured mesh, where the cell size was halved at each level }closer to the sphere, as shown in figure~\ref{fig:NumMesh}. The resolution of the coarse mesh used for $Re=\{200,300,500\}$ was roughly 3 million cells while the fine mesh used for $Re=1000$ consisted of roughly 17 million cells. Velocity fields were sampled at a convective time increment of $\Delta t_{DNS} = R/U_{o}$ for $Re = \{200,300,500\}$ and $\Delta t_{DNS} = 0.4R/U_{o}$ for $Re = 1000$. The number of snapshots saved ranged between 100-1250 depending on the complexity of the flow.  However, in all cases snapshots were saved over a minimum of five dominant oscillation periods. For further computational details, the reader is referred to \citet{ohh2021wake}. 

For each of the $n_r = 24$ cases in the $(Re,Fr)$ parameter space, dynamic mode decomposition was carried out using data obtained over the region $x \in [20.25R, 29.75R]$, $y\in [-4R, 4R]$, and $z \in [-4R, 4R]$ with a uniform grid spacing of $(1/8)R$ in each direction. 
This region was chosen as the observation window since it is still within the near-wake but does not include developments immediately behind the sphere (i.e., the recirculation zone). DMD was carried out using the method described in \citet{chen2012variants} from the time-resolved sequence of $S+1$ velocity field snapshots from DNS, $\{\mathbf{u}_0,\mathbf{u}_1,\cdots,\mathbf{u}_S\}$, which generated a set of DMD spatial modes (or Ritz vectors) $\{\mathbf{v}_1, \mathbf{v}_2,\cdots,\mathbf{v}_S\}$ and the corresponding temporal coefficients (or Ritz values) $\{\lambda_1,\lambda_2,\cdots,\lambda_S\}$.
Since DMD is an established technique, the details of this procedure are not repeated here. For the lower Reynolds number cases ($Re= \{200, 300, 500\}$), DMD was carried out using a sequence of $S+1 = 40$ snapshots from DNS obtained at time intervals of $\Delta t_{DNS} = R/U_{o}$.  Thus, the total duration of the sequence was $39(R/U_{o})$, which is roughly 4 flow through times for the streamwise extent of the snapshots used for DMD ($9.5R$).  The use of 40 snapshots generated $S=39$ DMD modes for each case. The DNS time step for $Re = 1000$ cases was lower, and the velocity snapshots were available at $\Delta t_{DNS} = 0.4R/U_{o}$. Therefore we used a total of $S+1=100$ snapshots for DMD to keep the total duration constant at roughly 4 flow through times. This resulted in 99 DMD modes for each of $Re=1000$ cases. It is important to note that $\Delta t_{DNS}$ is the time difference between consecutive snapshots in the DNS database and not the time step used in the DNS. For each $Re-Fr$ combination, the first 35 modes sorted in decreasing order of energy were retained and used to construct a library (or database) of DMD modes for sparse regression. We used an equal number of modes for each case to eliminate any regression bias due to inclusion of more modes from a particular parameter combination.

Note that the time-evolution of the snapshots can be expressed in terms of the complex Ritz vectors and values as:
\begin{subequations}
\begin{equation}
\mathbf{u}_{k} = \sum_{j=1}^{S} \lambda_{j}^{k} \mathbf{v}_{j}, \hspace{16pt} k = 0,\cdots, S-1.
\end{equation}
\begin{equation}
\mathbf{u}_{S} = \sum_{j=1}^{S} \lambda_{j}^{S} \mathbf{v}_{j} + \mathbf{r}, \hspace{16pt} \mathbf{r} \perp \text{span}\{\mathbf{u}_{1}, \mathbf{u}_{2}, \cdots, \mathbf{u}_{S-1}\}
\end{equation}
\end{subequations}

where $\mathbf{r}$ is the resulting residual for the last snapshot. Since the Ritz values determine the time evolution from one snapshot to the next, the Strouhal number for the $j^{th}$ DMD mode is given by 
\begin{equation}
St_{j} := \frac{\omega_{j}R}{2\pi U_{o}} = \frac{\mathrm{arg}(\lambda_{j})R}{2\pi \Delta t_{DNS} U_{o}}
\end{equation}
where $\Delta t_{DNS}$ is the time interval between two consecutive snapshots in the DNS database and $\omega_j$ represents the oscillation frequency. Note that the snapshots used here are the three-dimensional (3D) flow fields from DNS and so the spatial modes obtained via DMD are also 3D flow fields with the same spatial extent. These modes must therefore be sub-sampled to yield 2D snapshots or point measurements when the DMD mode database is compiled for regression and regime identification. 
Further details on how the database is compiled for different input data are provided below in Sec.\ref{sec:prob_form}. 

\subsection{Laboratory Experiments}\label{sec:experiments}

\begin{figure}
 \center
 \includegraphics[width=0.75\linewidth]{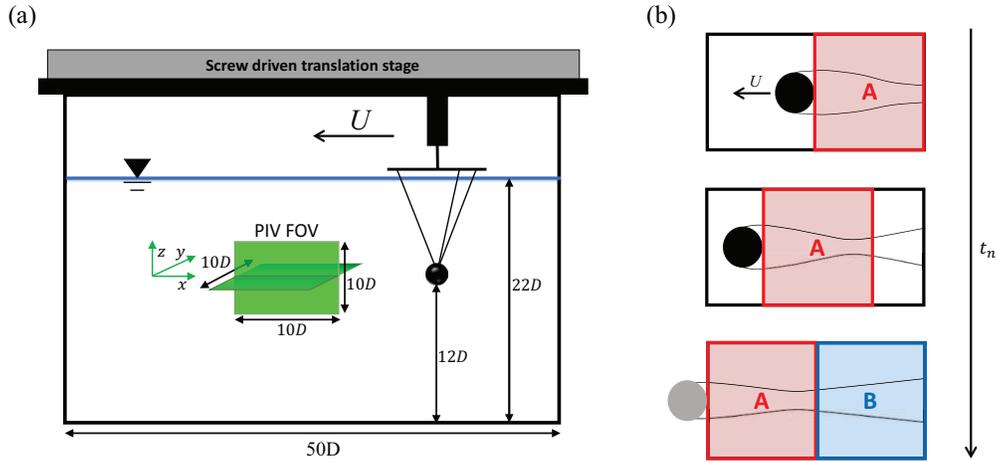}
 \caption{\label{fig:ExpDetail}(a) Schematic showing experimental setup. The annotated scale is based on a typical sphere diameter of $D= 4$ cm. (b) Schematic showing temporal data reconfiguration where the colored box $A$ is moving in reference to the sphere.}
 \end{figure}

Experiments were carried out at values of $Re$ and $Fr$ that match the simulations (figure~\ref{fig:ReFrspace}). 
 A sphere was towed horizontally in a tank of size $1 \mathrm{m} \times 1 \mathrm{m} \times 2.5 \mathrm{m}$, filled with stably stratified fluid as illustrated in figure~\ref{fig:ExpDetail}(a). To achieve a linear density gradient while avoiding optical distortions, a salt and ethanol water mixture was used to match refractive index \citep{xiang:15}. The sphere consisted of a 3D-printed polylactic acid (PLA) shell around a steel ball.  This sphere was submerged to the center of the tank in the lateral and vertical directions with three fishing wires $(d=0.5 mm)$. To adjust $Re$ and $Fr$, the buoyancy frequency, sphere radius and tow speed were varied systematically over the following ranges: $N \in [0.13,1.0]$ rad/s, $R \in [0.72,5.5]$ cm, and $U_{o} = [0.37-43]$ cm/s.

Planar velocity measurements were obtained in the wake of the sphere using Particle Image Velocimetry (PIV). The tank was seeded with titanium dioxide particles of average density $4.23$ g/$\mathrm{{cm^3}}$ and diameter $15$ $\mathrm{\mu m}$. The vertical and horizontal center planes were illuminated with a pulsed laser (Nd:YAG, LaVision NANO L100-50PIV). Two cameras (LaVision-Imager sCMOS), each having a resolution of $2560 \times 2160$ pixels, were aligned side-by-side with $\sim 5 \%$ overlap to obtain an extended field of view (FOV) in the streamwise direction. For measurements in the vertical ($x-z$) plane, the FOV for each camera was approximately 350 mm $\times$ 300 mm.  Given the varying sphere sizes, this translates into a scaled FOV that varies from $6.5R \times 5.5R$ to $49R \times 42R$. For measurements in the horizontal ($x-y$) plane, the FOV for each camera was roughly 230 mm $\times$ 190 mm, which translates into a scaled field of view ranging from $4.2R \times 3.5R$ to $32R \times 26R$. Images were taken at a single-pulsed time interval ranging between $\Delta t_{PIV} = 0.05-0.2$ s, which was chosen to ensure that the particle pixel displacements between images produced a well-resolved velocity field. 

To ensure consistency with the numerical simulations, the experimental data were translated to a sphere-fixed frame. A fixed sphere reference frame was achieved by stacking half the streamwise span of the FOV, marked $A$ in figure~\ref{fig:ExpDetail}(b), into a temporal sequence. 
Note that, box A moved with the sphere as it is towed across the FOV. Once box A reaches the upstream edge of the FOV, the next half of the FOV (marked B in figure~\ref{fig:ExpDetail}(b)) becomes available. The duration of the time series available in the fixed-sphere reference frame was restricted to the period in which box A remained within the FOV. For the experiments, this ranged between $2.2-21.6$ s and the number of snapshots available ranged between $m = x_{\mathrm{FOV}}/(2U_{o}\Delta t_{PIV}) = 60-300$ frames.  Where possible, velocity measurements were extracted over a streamwise extent corresponding roughly to the DMD domain from these re-sequenced datasets.  However, it was not always possible to ensure an exact overlap with the DMD domain of $x \in [20.25R,29.75R]$. Indeed, for several low Froude number cases, the streamwise window available from PIV measurements was significantly smaller (approximately $3.5R$) than the DMD domain.

\subsection{Problem formulation}\label{sec:prob_form}
Briefly, the regime identification problem can be stated as follows.  Given a time series of input data from an unknown regime and a large database of DMD modes from known regimes, we seek a sparse representation of the training data in terms of the DMD modes in the database. In other words, a model is constructed for the input flow field, approximating it as a linear combination of the dominant DMD modes identified from the large database. The parameter regime is then estimated by taking a projection-weighted average of the known $Re$ and $Fr$ values for the selected DMD modes.  We consider regime identification using two different types of input data: (a) planar (two-dimensional, two-component; 2D2C) velocity fields or (b) point measurements of all three-components of velocity from a limited set of sensors. The problem formulation is similar in both scenarios, though the DMD modes must be sub-sampled differently depending on the input data prior to forming the database.

\begin{figure}
    \centering
    \includegraphics{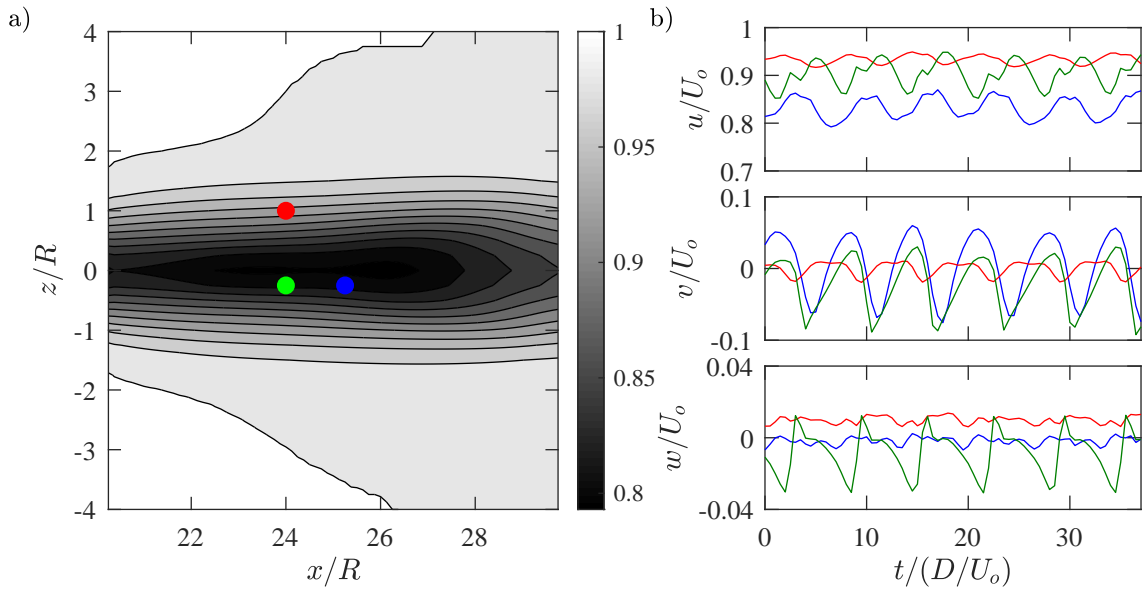}
    \caption{(a) Streamwise velocity contours normalized by freestream velocity $U_{o}$. Three point sensors are shown in blue, green, and red. These sensors are displaced from a common central location, $(x,y,z) = (24R,-0.25R,-0.25R)$, by a distance of $1.25R$ in the streamwise, lateral, and vertical directions, respectively (\textit{n.b.}, the green sensor is not in the same $x-z$ plane as the blue and red sensors). (b) Streamwise, vertical, and spanwise velocity signals are plotted against time for all the three sensors. The signals are color coded by sensor.}
    \label{fig:TRschematic}
\end{figure}

The 2D2C snapshots used for regime identification are obtained in the central $x-z$ (streamwise-vertical) plane of the sphere, i.e., corresponding to $y=0$. If 2D2C snapshots from DNS are used, these include data over the same streamwise and vertical extents as the DMD modes.  If 2D2C snapshots from experiments are used, the streamwise extent varies slightly from case to case but is comparable to that for the DMD modes.  The point measurements used for regime identification are extracted from the DNS database. In this case, we assume the availability of 3 sensors, each capable of measuring all components of velocity. These 3 sensors are offset by a distance of $1.25R$ in streamwise, lateral, and vertical directions, respectively, from a common central point, which we term the \textit{sensor location}.  This setup is shown in figure~\ref{fig:TRschematic}. 

The time series of $K$ velocity measurements (or inputs) obtained at intervals of $\Delta t$, $\{\mathbf{z}_1,\mathbf{z}_2,\cdots,\mathbf{z}_K\}$, is first stacked into a single column vector of the form: 
\begin{equation}
\mathbf{z} = 
\begin{bmatrix}
\mathbf{z}_{1} \\
\mathbf{z}_{2} \\
\vdots \\
\mathbf{z}_{K}
\end{bmatrix}.
\end{equation}
Note that each individual element of $\mathbf{z}$ represents a complete 2D2C spatial snapshot or all the velocity readings from the point sensors. Since these input measurements are obtained at different spatial and temporal resolutions relative to the DMD dataset, the spatial DMD modes must be sub-sampled and the time evolution must be modified before the database of DMD modes is created.  To ensure compatibility with the input data, a time sequence from DMD mode $j$ in regime $i$ is created as follows:

\begin{equation}
\mathbf{p}_{j,i,h} = 
\begin{bmatrix}
\lambda_{j,i}^{0}\mathbf{S}_h \mathbf{v}_{j,i} \\
\lambda_{j,i}^{\Delta t^{*}}\mathbf{S}_h \mathbf{v}_{j,i} \\
\vdots \\
\lambda_{j,i}^{(K-1)\Delta t^{*}}\mathbf{S}_h\mathbf{v}_{j,i}
\end{bmatrix}
\end{equation}
where $\Delta t^{*}$ is the ratio of the sampling rates of training data and the DNS database used for DMD (i.e., $\Delta t^{*} = \Delta t / \Delta t_{DNS}$), $\lambda_{j,i}$ and $\mathbf{v}_{j,i}$ are the complex Ritz values and vectors for DMD mode $j$ from regime $i$, and $\mathbf{S}_h$ is the $h^{th}$ sub-sampling matrix that is spatially compatible to the input data. For the 2D2C input data, we simply extract the central plane from the DMD spatial modes and so only a single sub-sampling matrix is used. For the point sensor data, we extract all compatible 3-point combinations from the DMD spatial modes, i.e., we consider all potential values for the central sensor location. In this case, the regime identification algorithm also provides an estimate for the sensor location. Note that factor of $\Delta t^*$ in the exponential for the Ritz values ensures time evolution in the DMD modes at a rate consistent with the input data.

Next, we assume that the input data can be modeled as a linear combination of the compatible DMD mode sequences:
\begin{equation}\label{eq:lin_rel_piv}
\mathbf{z} \approx \Re \left( \sum_{j,i,h} c_{j,i,h}\mathbf{p}_{j,i,h} \right).
\end{equation}
Here, $c_{j,i,h} \in \mathbb{C}$ are complex coefficients that calibrate the amplitude and phase of each DMD mode sequence to match the training data, and $\Re(\cdot)$ represents the real component. Equation~\ref{eq:lin_rel_piv} can be rewritten as
\begin{equation}\label{eq:lin_rel_piv_real}
\mathbf{z} \approx \sum_{j,i,h} \left( \Re(c_{j,i,h}) \Re(\mathbf{p}_{j,i,h}) - \Im(c_{j,i,h})\Im(\mathbf{p}_{j,i,h}) \right)
\end{equation}
in which $\Im(\cdot)$ represents the imaginary component. 

A large database is now constructed by pooling all the DMD mode sequences from each regime.  To avoid computation with complex coefficients, the real and imaginary parts of each compatible DMD mode sequence, $\mathbf{p}_{j,i,h}$, are included as separate basis functions $\mathbf{d}_m$ in this database. Thus, equation~(\ref{eq:lin_rel_piv_real}) is now reformulated as a linear regression problem as follows:
\begin{equation}{\label{eq:lin_reg_velField}}
\underbrace{
\left[
\begin{array}{c}
\mathbf{z}_1 \\
\mathbf{z}_2 \\ 
\vdots \\
\mathbf{z}_K
\end{array}
\right]}_{input\hspace{4pt}data} \approx 
\underbrace{
\left[
\begin{array}{c c c c c c c}
\vline & \vline &  & \vline \\ \\
\mathbf{d}_{1} & \mathbf{d}_{2} & \cdots & \mathbf{d}_{M} \\ \\
\vline & \vline &  & \vline
\end{array}
\right]
}_{basis\hspace{4pt}functions}
\underbrace{
\left[
\begin{array}{c}
b_1 \\
b_2 \\
\vdots \\
b_{M}
\end{array}
\right]
}_{coefficients}
\end{equation}
or equivalently,
\begin{equation}\label{eq:lin_reg_simple}
    \mathbf{z} \approx  \mathbf{D} \mathbf{b}.
\end{equation}
Here, $\mathbf{D}$ represents the database of $M$ different DMD mode sequences (or basis functions) and $\mathbf{b}$ is the column vector of coefficients for each of these basis functions.
We seek a sparse solution in which only a small number of coefficients are non-zero. In other words, we seek the dominant DMD mode sequences that best approximate the input. 

Recall that a total of $n_r=24$ different $(Re,Fr)$ combinations were considered in DNS corresponding to 4 different $Re$ values and 6 different $Fr$ values. For the simulations at $Re=\{200,300,500\}$, 40 snapshots were used for DMD, yielding 39 DMD modes.  For the simulations at $Re=1000$, 100 snapshots were used for DMD, yield 99 DMD modes. However, only the first 35 modes from each case were used to construct the database.  Thus, there were a total of $35 \times 24 = 840$ different DMD modes available for the regression problem. For the 2D2C snapshots of velocity, each DMD mode yielded 2 different basis functions (real and imaginary part of $\mathbf{p}_{j,i,h}$), for a total of $M = 1680$ basis functions in the database $\mathbf{D}$. For the point measurement input data, there were 252 different possible sensor locations (i.e., sub-sampling matrices $\mathbf{S}_h$) spaced at a distance of $1.25R$ in the $x$, $y$, and $z$ directions across the 3D spatial modes obtained from DMD. As a result, each DMD mode yielded $252 \times 2 = 504$ potential basis functions, which led to $M = 840 \times 504 = 423360$ basis functions in the database. 

For regime identification with 2D2C snapshots from DNS, we obtain the streamwise and vertical components of velocity ($u$, $w$) over a grid of size $n_{x}\times n_{z} = 77\times 65$ in the streamwise and vertical directions. In general, we use 10 consecutive snapshots of the velocity field obtained at a sampling rate of $\Delta t = 2R/U_{o}$, such that $\Delta t^* = (\Delta t/\Delta t_{DNS}) = 2$ for $Re = \{200,300,500\}$ and $\Delta t^* = 5$ for $Re = 1000$. Note that these data do not coincide with the DNS time series used for DMD. For regime identification from experimental data, we use a comparable spatial resolution though the sampling rate, $\Delta t$, is different for each case. However, for each case, we ensure that the DMD modes in the database are compiled at the sampling rate corresponding to the input data.  
The input data are split into training, validation and testing sets in the ratio of $60\%:20\%:20\%$, i.e., 6 snapshots for training, 2 for validation, and 2 for testing. A model is constructed for the training data, the so-called hyper-parameters (in this case, the number of modes retained in the regression) are fine-tuned based on the validation data, and the prediction confidence is evaluated using the testing data. 

For regime identification using point measurements, 3 different sensor locations in the wake were considered. The first location was directly in the center of the wake. The second and third locations were offset in the vertical and spanwise directions, respectively. Specifics regarding sensor location are provided in Sec.~\ref{sec:TRpoint}. We assume each sensor measures all 3 velocity components at the sampling rate used for 2D2C snapshots, $\Delta t = 2R/U_{o}$. A time series of 40 measurements is used as the input for regime identification. As before, these data are further split into training, validation, and testing sets in the ratio of $60\%:20\%:20\%$, i.e., 24 point measurements for training, 8 for validation and 8 for testing.

\subsection{Identifying dominant DMD modes using FROLS}\label{sec:ident_res}

The vector of coefficients $\mathbf{b}$ in Eqs.~\ref{eq:lin_reg_velField} and~\ref{eq:lin_reg_simple} can be computed using the standard least squares method. However, this will result in overfitting due to the large number of modes used in the database. Therefore, we seek a sparse solution to this problem where the number of modes used to reconstruct the flow field ($M_{o}$) is much less than the number used in the database ($M$). The reconstructed flow field ($\hat{\mathbf{z}}$) can then be represented as:
\begin{equation}\label{eq:reconst}
\hat{\mathbf{z}} = b_{l_1} \mathbf{d}_{l_1} + b_{l_2}\mathbf{d}_{l_2} + \cdots + b_{l_{M_{o}}}\mathbf{d}_{l_{M_{o}}}
\end{equation}
in which $M_{o} << M$ and $l_{m}$ are the indices of the selected modes in the database. The above equation in matrix form is
\begin{equation}\label{eq:predictor}
\hat{\mathbf{z}} = 
\underbrace{
\left[
\begin{array}{c c c c c c c}
\vline & \vline &  & \vline \\ \\
\mathbf{d}_1 & \mathbf{d}_2 & \cdots & \mathbf{d}_{M} \\ \\
\vline & \vline &  & \vline
\end{array}
\right]
}_{basis\hspace{4pt}functions}
\underbrace{
\left[
\begin{array}{ c }
0 \\
\vdots \\
b_{l_1} \\
0 \\
\vdots \\
b_{l_{M_o}} \\
0	\\
\vdots
\end{array}
\right]
}_{sparse\hspace{4pt}coeffs.}
\end{equation}
Such problems are typically solved using an iterative method that sequentially reduces the complexity of the model \citep[see e.g.,][]{rudy2017data, brunton2016discovering}. These studies employ standard techniques such as lasso \citep{tibshirani1996regression} or ridge \citep{hoerl1970ridge} regression, which solve a least squares regression problem with regularization, sequentially eliminating coefficients below a certain threshold until the required degree of sparsity is achieved. However, these algorithms initially start with a complex (non-sparse) model and iteratively reduce complexity (or promote sparsity) until an optimal model is reached. If the database is huge, either due to the number or size of the basis functions, the computational cost incurred for iterating can be high. Instead, we use the Forward Regression with Orthogonal Least Squares (FROLS) algorithm, which starts with a simple model and increases complexity (or decreases sparsity) with the number of iterations. This algorithm has previously been used for nonlinear system identification problems \citep{billings2013nonlinear}. We propose to use a similar approach to identify the dominant DMD modes and also calibrate their amplitude and phase.


The FROLS algorithm is an iterative process with two steps per iteration: 1) Ranking and 2) Gram-Schmidt orthogonalization. At each step a full search of the unselected modes is carried out to pick the best candidate from a large database. Recall that $\mathbf{D}$ is the full database of basis functions, $\mathbf{D} = \{ \mathbf{d}_1, \mathbf{d}_2, ..., \mathbf{d}_M \}$, which is composed of $M$ candidate basis functions. We are looking for the subset of these basis functions, $\mathbf{D}_{M_{o}} = \{ \mathbf{d}_{l_1}, \mathbf{d}_{l_2}, ..., \mathbf{d}_{l_{M_o}}  \}$, that best approximates the training snapshots denoted $\mathbf{z}$ so that $M_o << M$. In the first ranking step, the error reduction ratio, $ERR$, for each basis function is computed as
\begin{equation}\label{eq:ERR}
 ERR := \left(\frac{\mathbf{z}^{T}\mathbf{d}_{m}}{\mathbf{d}_{m}^{T}\mathbf{d}_{m}}\right)^2 \frac{\mathbf{d}_{m}^{T}\mathbf{d}_{m}}{\mathbf{z}^{T}\mathbf{z}}= \frac{\left(\mathbf{z}^{T}\mathbf{d}_{m}\right)^2}{||\mathbf{z}||^2||\mathbf{d}_m||^2}
\end{equation}
which is equivalent to `$\cos^2(\theta)$' between the input data and each of the basis functions. The basis function with the highest \ml{$ERR$} is selected; this is denoted $\mathbf{o}_{1}$, and is identical to $\mathbf{d}_{l_1}$ in $\mathbf{D}$. This basis function is then added to the subset $\mathbf{D}_{M_o}$ and a new set of orthogonal basis functions, $\mathbf{O}$. In the next step, which is the Gram-Schmidt orthogonalization, the projection of each of the basis functions in $\mathbf{D}$ along $\mathbf{d}_{l_1}$ is subtracted to create a modified database $\hat{\mathbf{D}}_1$. The process is then repeated with $\hat{\mathbf{D}}_1$ to identify the next orthogonal basis function $\mathbf{o}_2$. This continues until a satisfactory solution to the regression problem is achieved.

\begin{figure}
 \centering
 \begin{subfigure}[b]{0.45\linewidth}
 \centering
 \includegraphics[trim = {4.15cm 3cm 0cm 1.5cm},scale=0.65,clip=true]{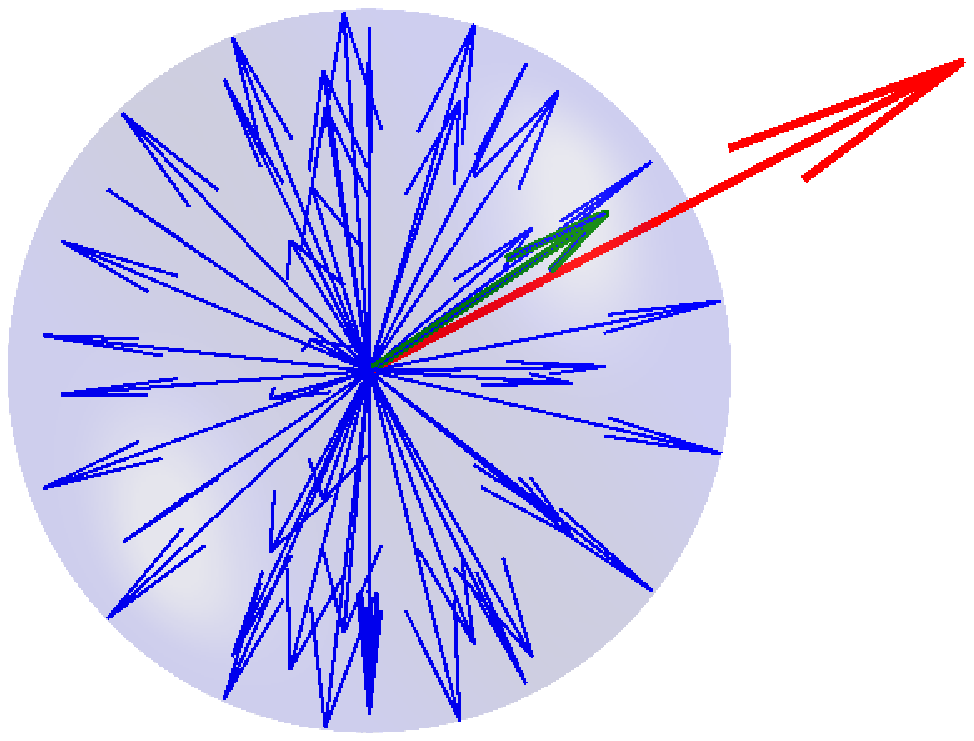}
 \caption{ }
 \end{subfigure}
 \begin{subfigure}[b]{0.45\linewidth}
 \centering
 \includegraphics[trim = {4.5cm 3cm 3cm 1.5cm},scale=0.65,clip=true]{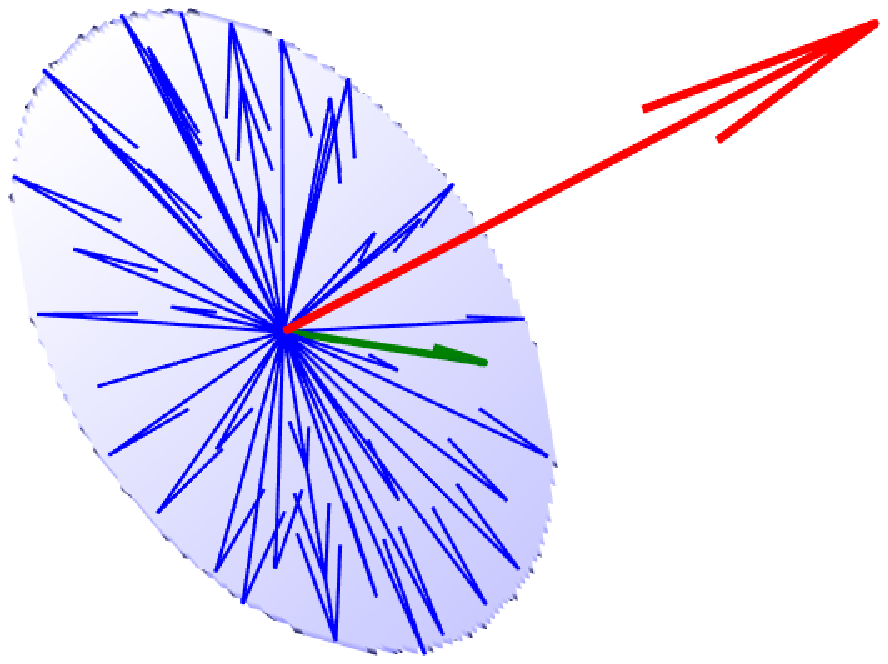}
 \caption{ }
 \end{subfigure}
 \caption{Schematic showing successive steps in the FROLS algorithm. Subplot (a) shows the database of basis functions in blue. The training vector (input data) is shown in red and the closest match to the training vector is shown in green. Subplot (b) shows the remaining basis functions after the projection along the green vector in (a) is subtracted.}
\label{fig:frols_schem}
 \end{figure}

The ranking and Gram-Schmidt orthogonalization steps are shown schematically in figure~\ref{fig:frols_schem}. Subplot (a) shows the training snapshot as the red vector and the (non-orthogonal) basis functions as blue vectors in three-dimensional space. The basis function \ml{with highest $ERR$} is shown in green. In the next step, the Gram-Schmidt Orthogonalization, this component is subtracted from rest of the vectors. Now the basis functions span a two dimensional space in contrast to three dimensions previously as shown in subplot (b). Again the basis function with highest $ERR$ is selected, shown as the green vector, and the process is repeated.

Thus, we start with $\mathbf{D} = \{\mathbf{d}_{1}, \mathbf{d}_{2}, \mathbf{d}_{3}, ..., \mathbf{d}_{M}\}$ and identify the orthogonal set  $\mathbf{O} = \{\mathbf{o}_{1}, \mathbf{o}_{2}, \mathbf{o}_{3}, ... \mathbf{o}_{M_{o}}\}$ with $M_{o} << M$. Note that the orthogonal set $\mathbf{O}$ and $\mathbf{D}_{M_o}$ are both subsets of $\mathbf{D}$ and span the same space. The predictor can be represented in terms of either $\mathbf{O}$ or $\mathbf{D}_{M_o}$ as
\begin{equation}\label{eq:sparse_model}
\hat{\mathbf{z}} = \sum_{i=1}^{M_{o}}f_{i}\mathbf{o}_{i} = \sum_{i=1}^{M_{o}}b_{l_i}\mathbf{d}_{l_i}
\end{equation}
in which the coefficients for the orthogonal basis functions, $\mathbf{f}=\{f_1, f_2, \cdots, f_{M_o}\}$, can be computed using the orthogonality condition:
\begin{equation}
   f_{i} = \frac{\mathbf{z}^{T}\mathbf{o}_{i}}{\mathbf{o}_{i}^{T}\mathbf{o}_{i}}
\end{equation}
Each of the basis functions in $\mathbf{O}$ can be represented as a linear combination of basis functions in $\mathbf{D}_{M_{o}}$. In the next step, we map back the elements of $\mathbf{O}$ to $\mathbf{D}_{M_{o}}$. Note that the coefficients $\mathbf{b}_l = \{b_{l_1}, b_{l_2},\cdots,b_{l_{M_o}}\}$ can be computed from using the triangle equation
\begin{equation}\label{eq:triangle}
\mathbf{A}\mathbf{b}_l = \mathbf{f},
\end{equation}
where the upper triangular matrix
\begin{equation}\label{eq:triangle_A}
\mathbf{A} = 
\begin{bmatrix}
1 & a_{1,2} &  \cdots & a_{1,M_{o}} \\
0 & 1 & \cdots & a_{2,M_{o}} \\
\vdots & \vdots & \ddots & \vdots \\		 
0 & 0 & \cdots &  1
\end{bmatrix}
\end{equation}
is composed of elements $a_{r,s} = \mathbf{o}_{r}^{T}\mathbf{d}_{l_s} / \mathbf{o}_{r}^{T}\mathbf{o}_{r}$ and $a_{s,s} = 1$. 
The convergence can be assessed in terms of the reconstruction accuracy for the input data
\begin{equation}\label{eq:error}
\epsilon = \frac{\|\mathbf{z} - \hat{\mathbf{z}}\|}{\|\mathbf{z}\|}.
\end{equation}

The process described above is the standard FROLS algorithm mentioned in \citet{billings2013nonlinear}. Recall that the real and imaginary parts of the DMD modes are included in the database as separate basis functions (see equations~\ref{eq:lin_rel_piv_real} and \ref{eq:lin_reg_velField}). We implement a minor change to the algorithm in the ranking and selection step to select a complete DMD mode rather than just picking the real or imaginary part. This modification does not alter the underlying concept of the algorithm and is merely an adaptation for the problem at hand. This change is implemented as follows. After ranking each of the basis functions in the database, we select the basis function that has the highest \ml{$ERR$ (Eq.~\ref{eq:ERR})} with the training data and add it to $\mathbf{O}$. Then the rest of the basis functions in $\mathbf{D}$ are orthogonalized. Now, instead of ranking the modes in the database again, we pick the complex conjugate of the mode selected in the ranking step. For example, if the selected basis function is the real part of a particular DMD mode, we select its imaginary part and add it to $\mathbf{O}$. The ranking step is then repeated after orthogonalizing the rest of the basis functions in $\mathbf{D}$. This process ensures that both the real and imaginary parts of a particular DMD mode are included in the set $\mathbf{O}$. Note that inclusion of both the real and imaginary components of the DMD modes ensures that arbitrary phase shifts can be captured in the reconstruction.  This subtlety does not apply for DMD modes corresponding to the mean flow field.

\subsection{Estimating regime and quantifying accuracy and confidence}\label{sec:reconst}
Recall that the input data is split into three blocks: a training set, a validation set, and a testing set. The training data is used to construct the model, the validation set is used to fine-tune the hyperparameters (i.e., the number of FROLS iterations, $M_o$), and the testing set is used to evaluate the confidence in the predicted regime. To evaluate the optimum number of iterations, the error is tracked for the training and validation sets using Eq.~\ref{eq:error}. For the training set, we expect the error to decrease monotonically with the number of iterations. For the validation set, we expect the error to decrease initially but then increase once the model begins to \textit{overfit} to the training data. The iteration that yields the lowest error for the validation set is therefore deemed optimal and the calibration coefficients are evaluated using Eq.~\ref{eq:triangle} for this value of $M_o$. From the identified and calibrated modes in the database, the $(Re,Fr)$ regime of the input signal (together with location for the point measurements) is then estimated by computing a projection-weighted average

\begin{equation}\label{eq:regime_estimate}
\begin{split}
&\overline{Re}_{w} := \frac{\sum_{i = 1}^{M_{o}} W_{i}Re_{l_{i}}}{\sum_{i = 1}^{M_{o}} W_{i}}, \\
&\overline{Fr}_{w} := 10^{\left(\frac{\sum_{i = 1}^{M_{o}} W_{i}\log_{10}Fr_{l_{i}}}{ \sum_{i = 1}^{M_{o}} W_{i}}\right)},
\end{split}
\end{equation}
in which $Re_{l_{i}}$ and $Fr_{l_{i}}$ are the Reynolds number and Froude number for the selected basis function $l_{i}$. Note that the weighted average for $Fr$ is evaluated in logarithmic space rather than linear space to reflect the available DNS data, which spanned $Fr = 0.5$ to $Fr = 16$.
The scalar projection of the calibrated mode on to the training data is used as the weight:
\begin{equation}
W_{i} = \frac{(b_{l_{i}}\mathbf{d}_{l_{i}})\cdot\mathbf{z}}{\|\mathbf{z}\|^2}
\end{equation}
For regime identification using point measurements, the sensor location is also estimated using a projection weighted average of the sensor coordinates for the identified basis functions.

To quantify prediction accuracy over the parameter range covered by the DNS and experiments (see figure~\ref{fig:ReFrspace}), we transform the $Re$ and $Fr$ space into a unit square using the following transformation:
\begin{equation}
\begin{split}
&Re^{\prime} = \frac{Re - 200}{1000 - 200}, \\
&Fr^{\prime} = \frac{\log_{10}Fr - \log_{10}0.5}{\log_{10}16 - \log_{10}0.5}. 
\end{split}
\end{equation}
We then compute the distance between the prediction and the actual value and normalize it with the diagonal of the unit square. Thus, accuracy is defined as: 
\newcommand{\norm}[1]{\left\lVert#1\right\rVert}

\begin{equation}\label{eq:accuracy}
    \text{Accuracy: \quad} \alpha = 1 - \norm{
    \begin{bmatrix}
    Re^{\prime} - \overline{Re}_{w}^{\prime} \\
    Fr^{\prime} - \overline{Fr}_{w}^{\prime}
    \end{bmatrix}}/\sqrt{2}
\end{equation}

Since we use a weighted average, the worst possible prediction cannot exceed the diagonal of the unit square. This choice of normalization is to ensure that the error lies between 0 and 1. An error of 0 implies perfect prediction and an error of 1 implies the worst possible prediction.

\begin{figure}
    \centering
    \includegraphics[scale=0.75]{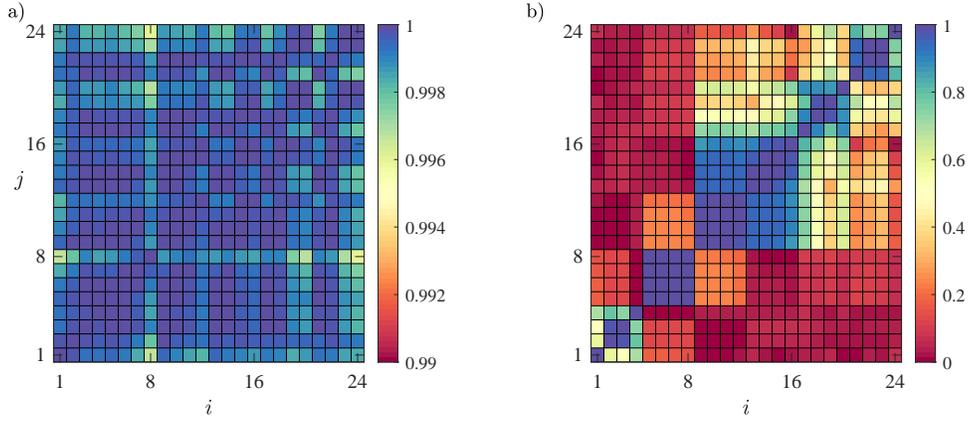}
    \caption{\ml{Component-wise} alignment between mean flow fields is evaluated between each of the $(Re,Fr)$ cases in figure~\ref{fig:ReFrspace}. There are a total of 24 cases, yielding a $24 \times 24$ grid. (a) Alignment between streamwise mean velocity fields, $| \overline{u}_i \cdot \overline{u}_j | / (\| \overline{u}_i \| \| \overline{u}_j \|)$, and (b) Alignment between vertical mean velocity fields, $| \overline{w}_i  \cdot \overline{w}_j | / (\| \overline{w}_i \| \| \overline{w}_j \|)$.  Here, the $(\cdot$) notation implies an element-wise product and sum.}
    \label{fig:meanAlign}
\end{figure}

Confidence in the constructed model, and hence the estimated regime, is evaluated using the testing data. The confidence is defined as the \ml{\textit{alignment}} between the predictor ($\hat{\mathbf{z}}$) and the input data ($\mathbf{z}$) over the testing data:
\begin{equation}\label{eq:conf}
    \text{Confidence: \quad} \chi = \left| \frac{\mathbf{z}_{test} \cdot \hat{\mathbf{z}}_{test}}{\|\mathbf{z}_{test}\| \|\hat{\mathbf{z}}_{test}\|}\right|,
\end{equation}
where the subscript is used to indicate that this metric is computed only over the testing data.  Further, we only compute confidence using the vertical velocity component ($w$) for reconstructions based on 2D2C velocity snapshots, and using both vertical and lateral components ($v$ and $w$) for reconstructions based on point measurements.  The streamwise component ($u$) is not included in the calculation because there is a very high degree of alignment between the mean streamwise fields ($\overline{u}$) of all the regimes, which dominates the confidence metric. Figure~\ref{fig:meanAlign} shows the alignment in mean flow fields between each of the different $(Re,Fr)$ cases for which we have DNS data.   There are a total of 24 cases and so we have a $24\times 24$ grid showing alignment between cases. As expected, the plot is symmetric and all the diagonal cases ($i=j$) show perfect alignment. Figure~\ref{fig:meanAlign}(a) shows that there is a very high degree of alignment between the streamwise mean velocity fields . In contrast, figure~\ref{fig:meanAlign}(b) shows that, for the mean vertical velocity, there is very little alignment between cases. Therefore, if the streamwise velocity is included in the computation of the confidence metric in Eq.~\ref{eq:conf}, then the reconstructions show high confidence levels irrespective of whether the model has the correct modes. Thus, using $w$ (and $v$) to evaluate confidence is more appropriate for the reconstructions.  Note that we also pursue reconstructions in which the mean flow field is removed entirely from the input data.  This is discussed in greater detail below. We recognize that prediction accuracy and confidence can be evaluated using metrics other than the ones defined in Eq.~\ref{eq:accuracy}-\ref{eq:conf}.  Nevertheless, we believe that these metrics provide useful insight into performance. 

\section{Results and Discussion}\label{sec:results}
In this section we discuss the results for various types of input data. Section~\ref{sec:2D2C-DNS} deals with cases in which 2D2C Velocity fields from numerical simulations are used as the input. We first evaluate the accuracy of the method described in the previous section using input data from a known regime and quantify the accuracy and confidence over the available $(Re,Fr)$ space. We then attempt regime identification with a sub-sampled database. Specifically, we include DMD modes from half the available cases in the database and again evaluate the regime predictions to test if the method is able to interpolate between regimes. In Section~\ref{sec:2D2C-EXP}, we use 2D2C PIV snapshots from experiments and the same database as before, i.e., DMD modes obtained via DNS. The experimental data is inherently noisy.  In addition, the spatial windows over which velocity fields are available do not always overlap with the spatial windows used for the DMD modes in the database. The accuracy and confidence is again quantified for the experimental data.  We see the results obtained in this section as an indicator of regime identification performance for real-world measurements. Finally, in Section~\ref{sec:TRpoint} we make use of three point measurements of velocity as the input and evaluate predictions using both the full and the sub-sampled DMD databases. In contrast to the database used for reconstruction of 2D2C velocity fields, the database for this scenario contains many possible triplets of velocity data from each DMD mode, corresponding to different sensor locations, as separate basis functions. Therefore, in addition to identifying the regime, we also attempt to identify the sensor location. 

\ml{Regime estimation and flow field reconstruction is first carried out for $(Re,Fr) = (500,4)$.  We focus on this particular case because it falls at the intersection between three different regimes (see figure~\ref{fig:ReFrspace}) which makes reconstruction challenging.  Following this, regime estimation, accuracy, and confidence are evaluated for all 24 $(Re,Fr)$ combinations in figure~\ref{fig:ReFrspace}. Regime estimation for each case is performed using the full flow field as well as the mean-subtracted flow field.  Note that mean subtraction is performed only for the input data used for regime identification, i.e., the 2D2C snapshots or point measurements that constitute $\mathbf{z}$ in Eq.~(\ref{eq:lin_reg_simple}). The DMD modes used to create the database of candidate basis functions, $\mathbf{D}$ in Eq.~(\ref{eq:lin_reg_simple}), are computed using the full flow field from DNS.} 
A summary of results for the different scenarios considered is provided in Table~\ref{tab:summary}.

\begin{table}[ht]
    \centering
    {\renewcommand{\arraystretch}{1.25}
    \begin{tabular}{|c|c|c|c|c|c|c|c|c|}
    \hline
        \multicolumn{3}{|c}{} &
        \multicolumn{3}{|c|}{Full flow field} &
        \multicolumn{3}{c|}{Mean-subtracted} \\ \cline{1-9}
        Input data & Source & Database & $\alpha$ (\%) & $\chi$ (\%) & $\alpha$ ($Fr$) (\%) & $\alpha$ (\%) & $\chi$ (\%)& $\alpha$ ($Fr$) (\%) \\ \hline
        
        2D-2C & DNS & Full & 92.8 & 87.3 & 97 & 83.8 & 68.8 & 91.2 \\ \cline{3-9} 
        Snapshots (10)&& Sub-sampled& 88.1 & 84.4 & 93 & 77.4 & 50.5 & 87.1 \\ \hline
        
        2D-2C & DNS & Full & - & - &- & - & - & - \\ \cline{3-9} 
        Snapshots (30)&& Sub-sampled& 88.2 & 80.4 & 92 & 76.1 & 35 & 82.8 \\ \hline
        
        2D-2C & EXP & Full & 71.8 & 66.7 & 78.6 & 68.3 & 38.1 & 79 \\ \cline{3-9} 
        Snapshots (10)&& Sub-sampled&-&-&-&-&-&- \\ \hline
        
        2D-2C & EXP & Full & 73.2 & 73.6 & 79.7 & 70.1 & 34.5 & 76.5 \\ \cline{3-9} 
        Snapshots (30) && Sub-sampled&-&-&-&-&-&- \\ \hline
        
        2D-1C & EXP & Full & 74.4 & 70.3 & 86.2 & 69.2 & 38.6 & 81.2 \\ \cline{3-9} 
        Snapshots (10) && Sub-sampled&-&-&-&-&-&- \\ \hline
        
        2D-1C & EXP & Full & 73.9 & 79.4 & 83.8 & 67.3 & 36.5 & 76.7 \\ \cline{3-9} 
        Snapshots (30) && Sub-sampled&-&-&-&-&-&- \\ \hline
        
        Point Meas. & DNS & Full & 95.5 & 83.4 & 95.7 & 84.2 & 49.3 & 91.1 \\ \cline{3-9}
        Location 1 && Sub-sampled& 75.9 & 71.8 & 80 & 71.7 & 43.5 & 80.7 \\ \hline
        
        Point Meas. & DNS & Full & 67.9 & 80.6 & 80.6 & 76.8 & 63.5 & 79.9 \\ \cline{3-9}
        Location 2 && Sub-sampled& 66.6 & 74.4 & 78.5 & 70.2 & 51.7 & 76.5 \\ \hline
        
        Point Meas. & DNS & Full & 70.8 & 78.3 & 77.1 & 76.8 & 66.6 & 88.4 \\ \cline{3-9}
        Location 3 && Sub-sampled& 71.8 & 68.0 & 78.6 & 68.7 & 56.2 & 80.6 \\ \hline
        
    \end{tabular}}
    \caption{The accuracy ($\alpha$, Eq.~\ref{eq:accuracy}) and confidence ($\chi$, Eq.~\ref{eq:conf}) metrics, averaged over all 24 $(Re,Fr)$ cases, using the full flow field and the mean-subtracted flow field. Results are shown for different input data and database sizes. The numbers in parentheses (10 or 30) indicate the number of 2D snapshots used in the input dataset.  For the cases in which 10 snapshots are used, the accuracy and confidence metrics are averaged over 3 different realizations.}
    \label{tab:summary}
\end{table}

\subsection{2D2C PIV snapshots from DNS}\label{sec:2D2C-DNS}
\begin{figure}
    \centering
    \includegraphics[trim = {0cm 0cm 0cm 0cm},scale=0.9,clip=true]{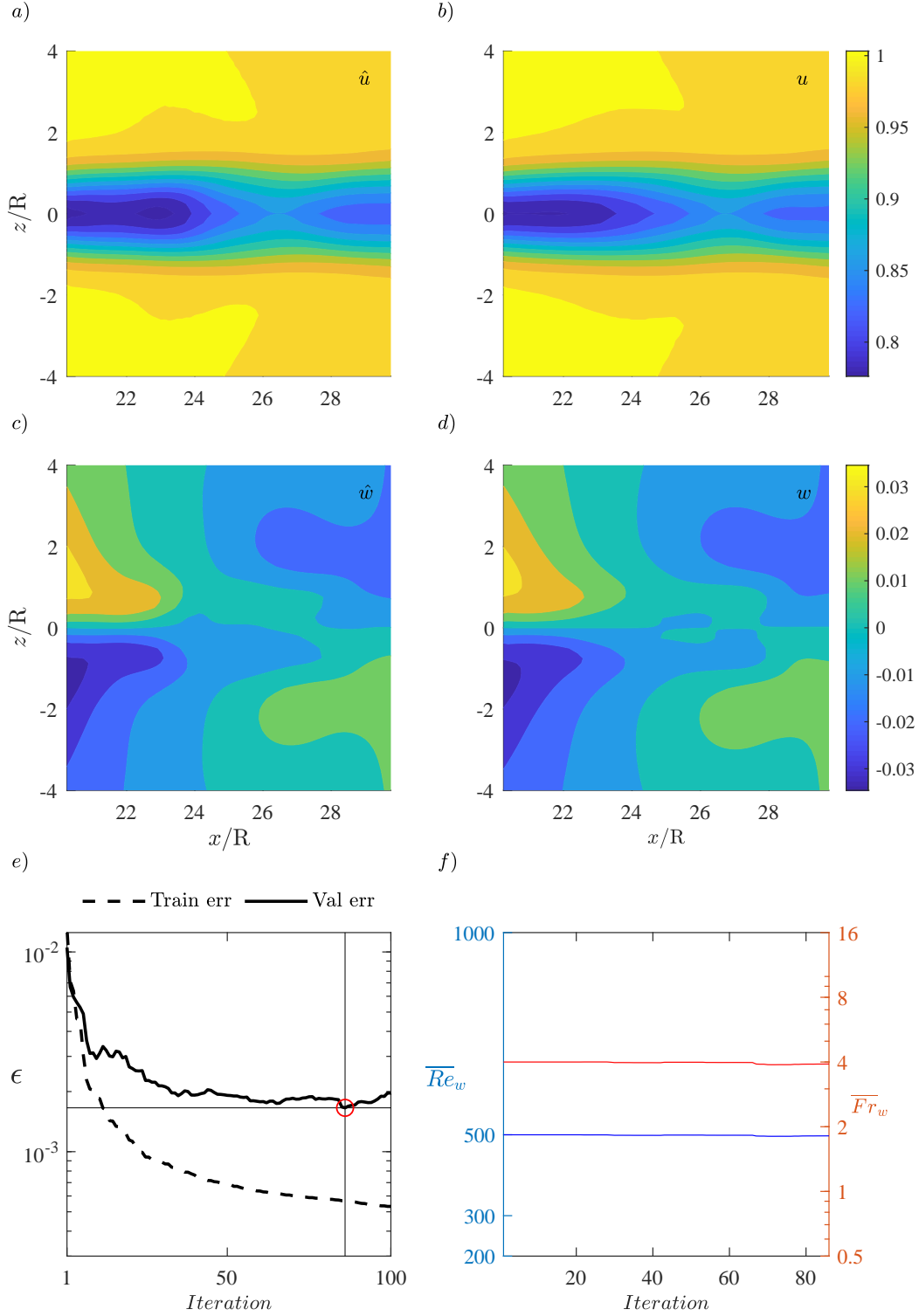}
    \caption{Comparisons between streamwise (a, b) and vertical velocities (c, d) over the testing data for $(Re,Fr) = (500,4)$. (e) Evolution of error with number of iterations for training and validation data sets. The minimum point for the validation error (i.e., the optimal iteration) is highlighted with a red circle. (f) The evolution of the predictions $\overline{Re}_{w}$ and $\overline{Fr}_{w}$ is shown until the optimal iteration.} 
    \label{fig:Re500Fr4MeanFullDbaseNumData}
\end{figure}

We first begin with input data from DNS for one of the 24 cases, $(Re,Fr)=(500,4)$, and quantify the accuracy of prediction and also assess the quality of the reconstructed flow field.  We use this specific parameter combination since it lies at the intersection of several different dynamic regimes (see figure~\ref{fig:ReFrspace}). For these reconstructions, we use a full database, i.e., with DMD modes from all 24 cases. The FROLS algorithm is then used to identify the DMD modes from this database that best represent the training data, and to calibrate their amplitude and phase. After calibration, the weighted DMD modes are used to reconstruct the flow field.  Importantly, note that the 10 snapshots used as the input data (6 for training, 2 for testing, 2 for validation) are different to the snapshots used for DMD. As expected, the method yields perfect predictions for $Re$ and $Fr$ if the snapshots used as input data are from the set used for DMD; these predictions are not shown here for brevity.

Figure~\ref{fig:Re500Fr4MeanFullDbaseNumData} shows comparisons between the DNS flow field and the reconstructed flow field for one of the test data snapshots. Panels (a) and (b) show comparisons for the streamwise velocity, while panels (c) and (d) show comparisons for the vertical velocity. The reconstructions are labelled with a $\hat{()}$. The reconstructed velocity field shows very good qualitative agreement with the velocity fields from DNS.
More quantitative results are shown in panel (e). The reconstruction error for the training data (Eq.~\ref{eq:error}) decreases monotonically whereas the error for the validation data reaches a minimum at the 86$^{th}$ iteration, which suggests that the regression algorithm begins to overfit to the training data beyond this point. Hence, we reconstruct the velocity field and identify the parameter regime using only the modes selected by the algorithm up to iteration 86. The corresponding regime prediction using the projection weighted average (Eq.~\ref{eq:regime_estimate}) is shown in panel (f).  Note that the algorithm yields a very close prediction, $\overline{Re}_w = 497$ and $\overline{Fr}_w = 3.9$. The confidence in these predictions, evaluated using Eq.~\ref{eq:conf}, is $99\%$.  Recall that the confidence is a measure of how similar the vertical velocity fluctuations in the reconstructed flow field are to the testing data. In this case, the vertical velocity fields for the testing data and the reconstruction look very similar.  As a result, we have a high level of confidence in the trained model and hence the predicted regime.

\begin{figure}
    \centering
    \includegraphics[scale=0.9,clip=true]{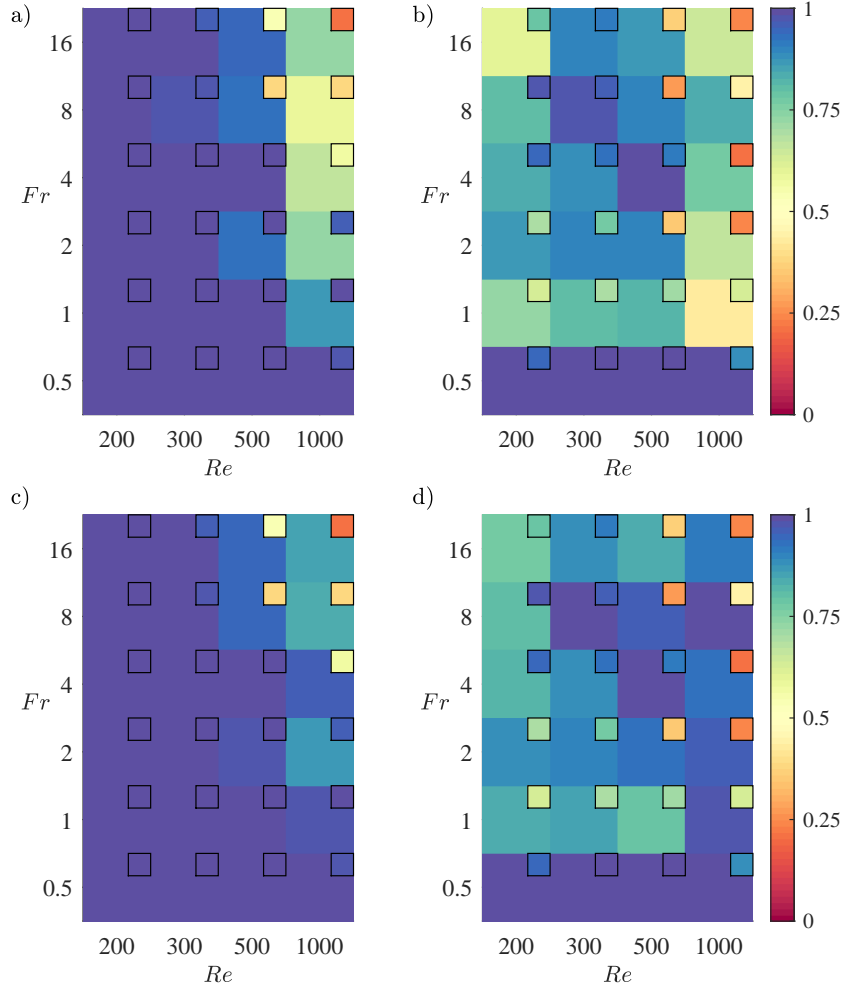}
    \caption{Plots showing prediction accuracy and confidence averaged over three different sets of input data for all $(Re,Fr)$ cases in the database. Regime predictions made using full flow fields as input data are shown in (a,c) and predictions generated using mean-subtracted flow fields are shown in (b,d). Large boxes represent prediction accuracy (Eq.~\ref{eq:accuracy}) while smaller inset boxes represent confidence (Eq.~\ref{eq:conf}). Panels (a,b) show the accuracy computed using both $Re$ and $Fr$, while panels (c,d) show the accuracy computed using $Fr$ alone.     
    }
    \label{fig:accuracy_FullDbase_ReFr}
\end{figure}

Next, we generate predictions for each of the 24 different $(Re,Fr)$ cases in the database. Figure~\ref{fig:accuracy_FullDbase_ReFr} shows the accuracy and the confidence of prediction computed using Eqs.~\ref{eq:accuracy} and~\ref{eq:conf}. The color of the larger square box represents accuracy while the color of the small inset box is an indicator for confidence. Here, we also compare regime predictions made using the full velocity field with predictions made using the mean-subtracted velocity field, i.e., using only the fluctuating velocity field. The accuracy and confidence metrics shown in figure~\ref{fig:accuracy_FullDbase_ReFr} are averaged over three different sets of input data. Overall, the prediction accuracy is higher when the mean flow is included in the training data, suggesting that the mean flow field contains useful information when it comes to distinguishing between $(Re,Fr)$ values. Specifically, the average accuracy across all 24 cases is approximately $93\%$ for regime predictions made using the full velocity field, compared to $84\%$ for the predictions made using the fluctuating velocity field alone (see Table~\ref{tab:summary}). \ml{A comparison of the results presented in figure~\ref{fig:accuracy_FullDbase_ReFr}(a) and figure~\ref{fig:accuracy_FullDbase_ReFr}(b) shows that, when the mean-subtracted velocity field is used, prediction accuracy deteriorates significantly for the cases that are classified as symmetric or asymmetric non-oscillation (red and yellow regions in figure~\ref{fig:ReFrspace}). This is expected since the dominant DMD modes for these regimes have $St = 0$ (figure~\ref{fig:velContours_regimes}). The FROLS algorithm is unlikely to select these $St = 0$ modes without mean flow information in the input data.} There is a qualitative correlation between prediction accuracy and the so-called confidence metric used in this study. The confidence metric is higher for cases in which the regime prediction is accurate, suggesting that the trained model is a useful future predictor of velocity, i.e., it yields accurate reconstructions for the testing data. 

Panels (a) and (b) in figure~\ref{fig:accuracy_FullDbase_ReFr} indicate that regime identification is less successful for cases with high Reynolds and Froude numbers.  Recall that the accuracy metric plotted in panels (a) and (b) represents the normalized distance between the actual $(Re,Fr)$ and the predicted values $(\overline{Re}_w, \overline{Fr}_w)$ over the parameter space considered in this study (Eq.~\ref{eq:accuracy}). Panels (c) and (d) in figure~\ref{fig:accuracy_FullDbase_ReFr} show accuracy computed using only the normalized distance between actual and predicted Froude number values, $Fr$ and $\overline{Fr}_w$.  A comparison between panels (a,b) and (c,d) shows that the framework developed here generates much better predictions for Froude number compared to Reynolds number.  For example, the average prediction accuracy for the high Reynolds and Froude number cases ($Re = 1000$ and $Fr \ge 2$) is $67\%$ when considering the normalized distance in $(Re,Fr)$ space compared to $88\%$ when only considering $Fr$. This indicates that the dominant source of error lies in the $Re$ prediction. One potential explanation for this is the fact that the high $Re$ and $Fr$ cases fall into a similar 3D dynamic regime (see figure~\ref{fig:ReFrspace}), and so the 2D2C snapshots used for regime identification and flow reconstruction may not have sufficient information to distinguish between different Reynolds numbers.  Another possible explanation is that the input sequence comprising 10 snapshots may not be sufficient to capture the wider range of timescales expected for the higher $Re$ cases. This is explored further below.

\begin{figure}
    \centering
    \includegraphics[scale=0.9,clip=true]{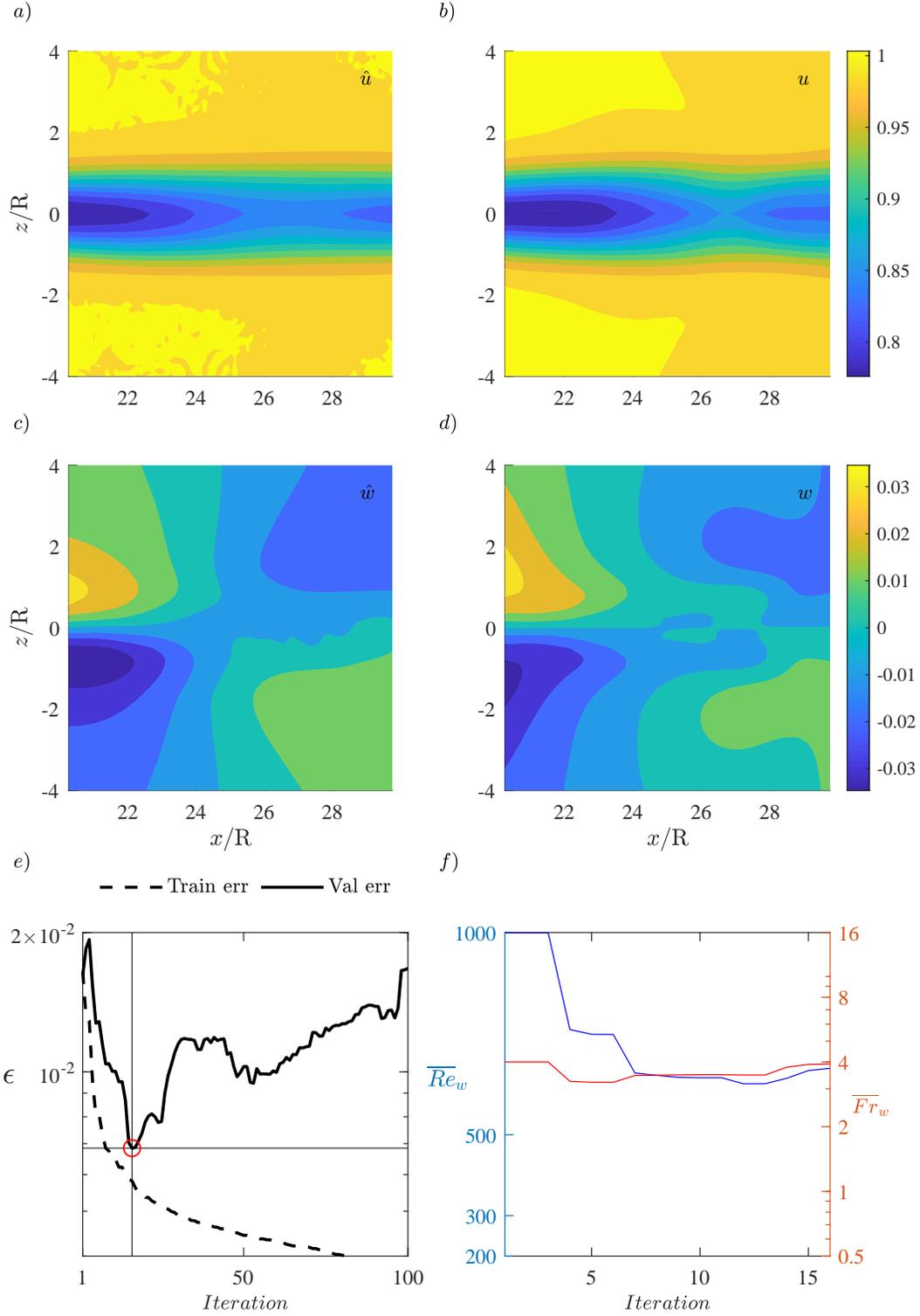}
    \caption{Comparison of reconstructions between streamwise (a, b) and vertical velocities (c, d) over the testing data for $(Re,Fr)=(500,4)$ using a sub-sampled DMD database. (e) Evolution of error with number of iterations for training and validation data sets. The minimum point for the validation error (i.e., the optimal iteration) is highlighted with a red circle. (f) The evolution in $\overline{Re}_w$ and $\overline{Fr}_w$ until the optimum iteration. The final prediction is $(\overline{Re}_{w}, \overline{Fr}_{w}) = (664, 3.9)$ with a confidence of $94\%$.}
    \label{fig:Re500Fr4MeanSparseDbaseNumData}
\end{figure}

As a further test, we evaluate prediction accuracy using a more limited database of DMD modes. In other words, we test whether the method is able to generate reasonable $(Re,Fr)$ predictions even if DMD modes from that specific parameter combination are not available in advance.  Specifically, we make use of a database with DMD modes from only half the available $(Re,Fr)$ cases; every other case in figure~\ref{fig:ReFrspace} is eliminated. We assess prediction accuracy for cases that are not in the database as well as the ones that are included. 

Figure~\ref{fig:Re500Fr4MeanSparseDbaseNumData} again shows predictions for input data from the case with $(Re,Fr)=(500,4)$.  However, these predictions are made using the more limited database, which did not contain DMD modes for this specific parameter combination. Panels (a)-(d) show that, even though the reconstructions are composed purely of DMD modes corresponding to other regimes, the reconstructed velocity fields qualitatively match the velocity fields from DNS for both the streamwise and vertical components. Interestingly, panel (f) shows that the FROLS algorithm picks a DMD mode corresponding to $(Re,Fr) = (1000,4)$ in the first iteration for this case. However, as the algorithm selects more modes the prediction evolves and eventually converges to a value of $(\overline{Re}_{w},\overline{Fr}_{w}) = (664,3.9)$, which is reasonably close to $(Re,Fr)=(500,4)$.  The final confidence level predicted using Eq.~\ref{eq:conf} is approximately $95\%$.

\begin{figure}
    \centering
    \includegraphics[trim = {0cm 7.5cm 0cm 0cm},scale=0.9,clip=true]{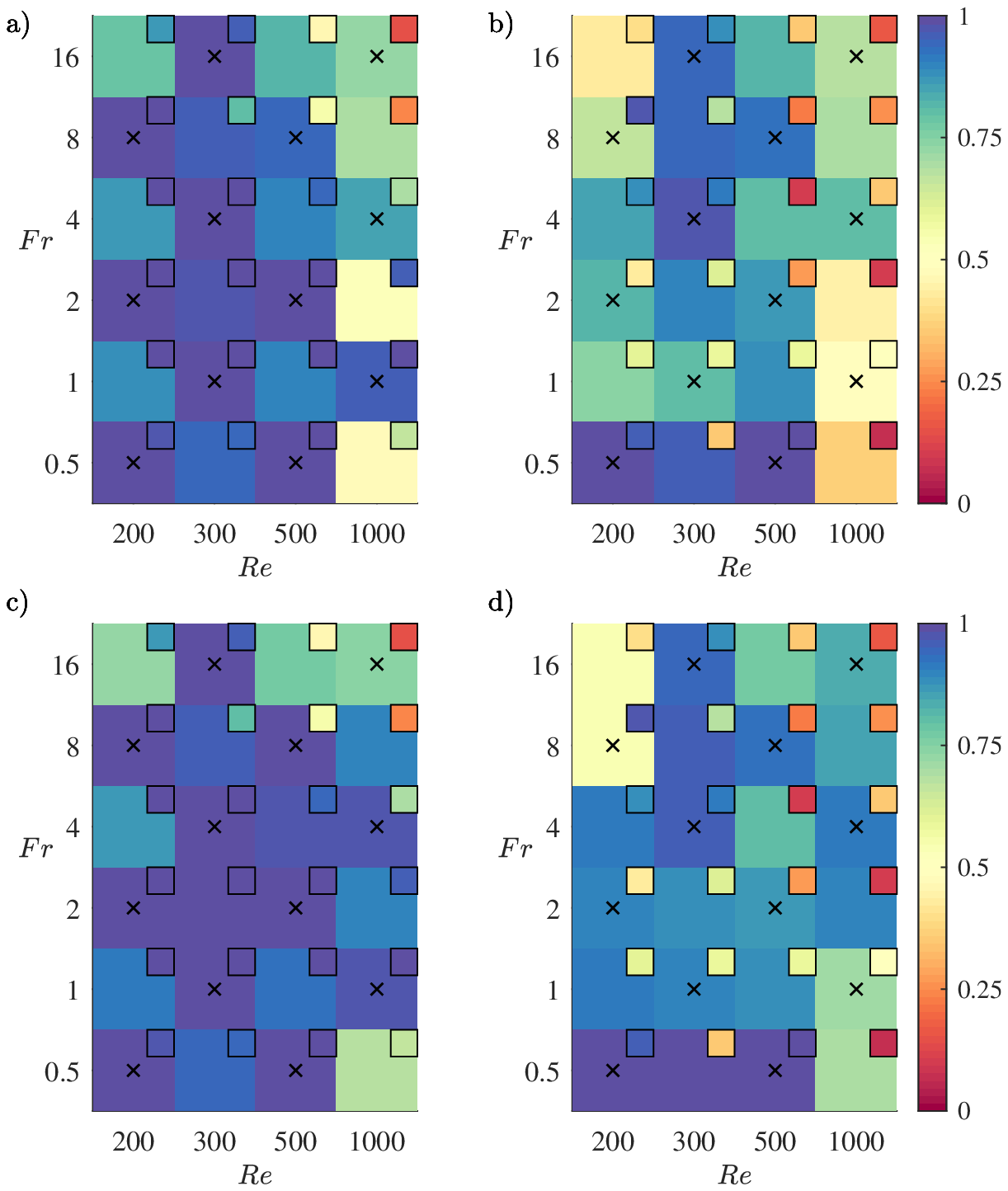}
    \caption{Plots showing prediction accuracy and confidence averaged over three different sets of input data.  These predictions are made using a database only consisting of DMD modes from the cases marked with a cross ($\times$). Regime predictions made using full flow fields as input data are shown in (a) and predictions generated using mean-subtracted flow fields are shown in (b). Large boxes represent prediction accuracy (Eq.~\ref{eq:accuracy}) while smaller inset boxes represent confidence (Eq.~\ref{eq:conf}).}
    \label{fig:accuracy_sparseDbase_ReFr}
\end{figure}

Figure~\ref{fig:accuracy_sparseDbase_ReFr} shows the accuracy and confidence metrics for all 24 $(Re,Fr)$ cases based on predictions made using the sub-sampled DMD database. Unsurprisingly, prediction accuracy and confidence are higher for input data from $(Re,Fr)$ combinations included in the database (marked with a $\times$).  For predictions made using the full flow field (figure~\ref{fig:accuracy_sparseDbase_ReFr}(a)), the average prediction accuracy is $95\%$ for $(Re,Fr)$ cases included in the database and $81\%$ for cases excluded from the database. As before, predictions made using mean-subtracted input data are less accurate than regime predictions made using the full velocity field (average accuracy is roughly $88\%$ for results shown in panel (a) and $77\%$ for results shown in panel (b)). Importantly, with the exception of a few outliers, prediction accuracy and confidence are reasonably high even for the $(Re,Fr)$ cases not included in the database.  This confirms that the method developed here is able to reconstruct flow fields from regimes not seen \textit{a priori}, as long as the database contains basis functions that can reasonably approximate the input data.  

Similar to the results shown in figure~\ref{fig:accuracy_FullDbase_ReFr}, the prediction accuracy is significantly higher for $Fr$ alone ($93\%$ on average for the full velocity field) compared to the accuracy for the combined $(Re,Fr)$ metric ($88\%$ on average), i.e., the $Re$ predictions are less accurate than the $Fr$ predictions. For brevity, these predictions are not shown here explicitly. Further, prediction accuracy is again lower for the cases with $Re = 1000$. As noted earlier, one possible explanation for this is that the short input sequence comprising 10 snapshots may not be sufficient to capture the more broadband temporal signature associated with wakes at higher $Re$. To evaluate whether a longer input sequence leads to an improvement in performance, we also attempted regime identification with 30 snapshots from DNS and the sub-sampled database of DMD modes. However, as shown in Table~\ref{tab:summary}, this did not yield significant changes in prediction accuracy or confidence. Table~\ref{tab:summary} provides a summary of prediction accuracy and confidence for the different scenarios tested.

\subsection{2D2C PIV snapshots from experiments}\label{sec:2D2C-EXP}

\begin{figure}
    \centering
    \includegraphics[scale=0.9,clip=true]{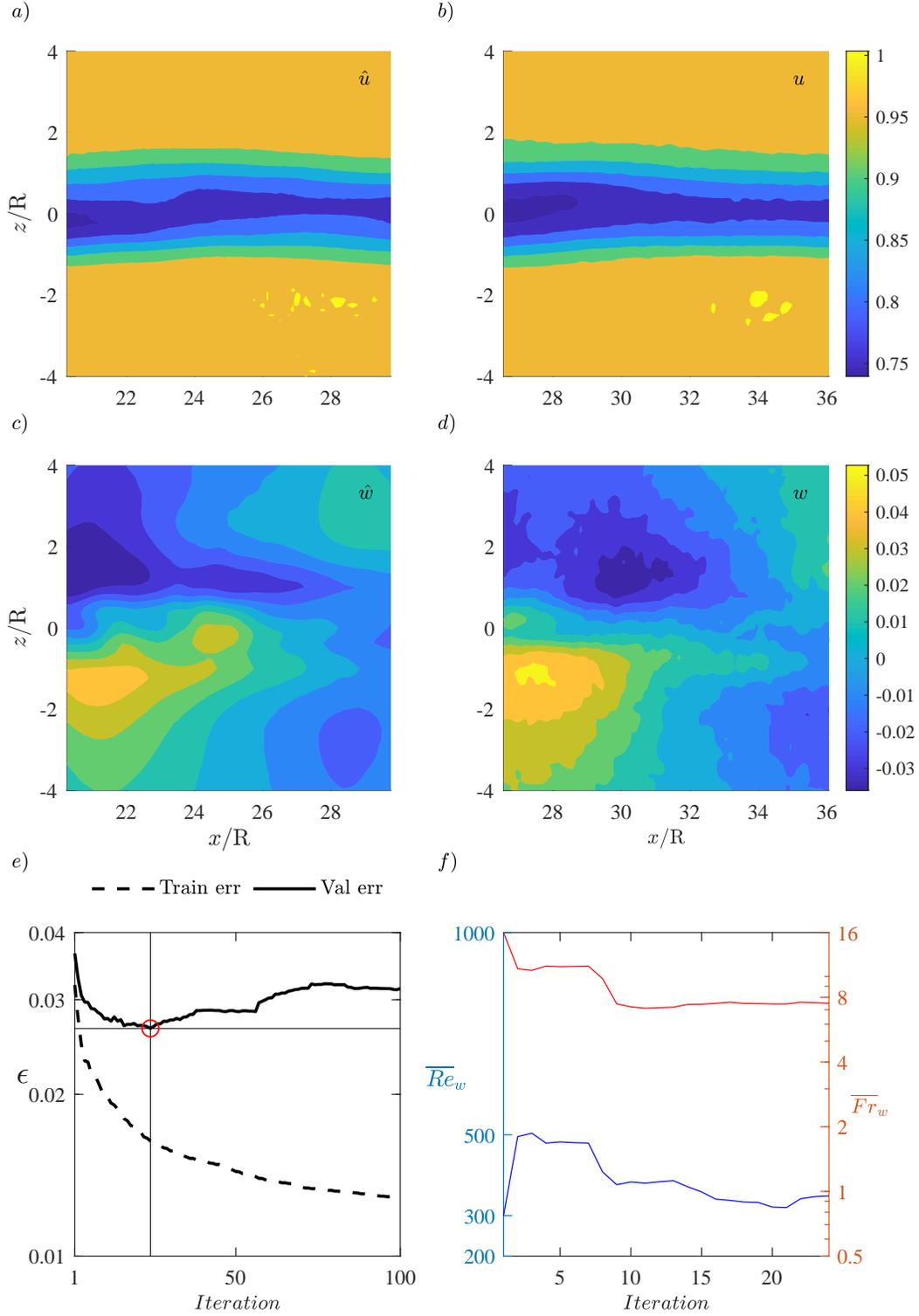}
    \caption{Comparisons between measurements and reconstructions for the streamwise (a, b) and vertical velocity (c, d) over the testing data for the experimental dataset at $(Re,Fr)=(533,3.7)$. Note the differing spatial extent of the wakes for the experimental and DNS data. (e) Evolution in error for the training and validation data sets. The minimum point for the validation error (i.e., the optimal iteration) is highlighted with a red circle. (f) The evolution of $\overline{Re}_w$ and $\overline{Fr}_w$ until the optimal iteration. The final prediction is $(\overline{Re}_{w}, \overline{Fr}_{w}) = (349, 7.5)$ with a confidence of 83\% per Eq.~\ref{eq:conf}.}
    \label{fig:Re500Fr4MeanFullDbaseExpData}
\end{figure}

\begin{figure}
    \centering
    \includegraphics[trim={0cm 7.5cm 0cm 0cm},scale=0.9,clip=true]{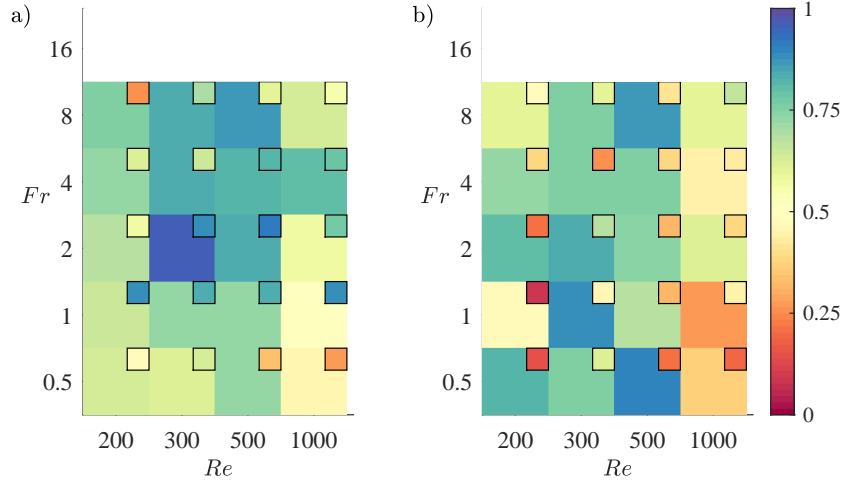}
    \caption{Prediction accuracy \ml{(large boxes, Eq.~\ref{eq:accuracy})} and confidence \ml{(smaller inset boxes, Eq.~\ref{eq:conf})} averaged over three snapshot sequences from experiments, using the full DMD mode database. As before, panel (a) shows results for the full flow field while panel (b) shows results obtained using the mean-subtracted flow field. Experimental data for the $Fr=16$ cases are unavailable. However, the corresponding DMD modes from DNS have been included in the database.}
    \label{fig:accuracyPlot_FullDbase_ExpData}
\end{figure}

\begin{figure}
    \centering
    \includegraphics[trim={0cm 7.5cm 0cm 0cm},scale=0.9,clip=true]{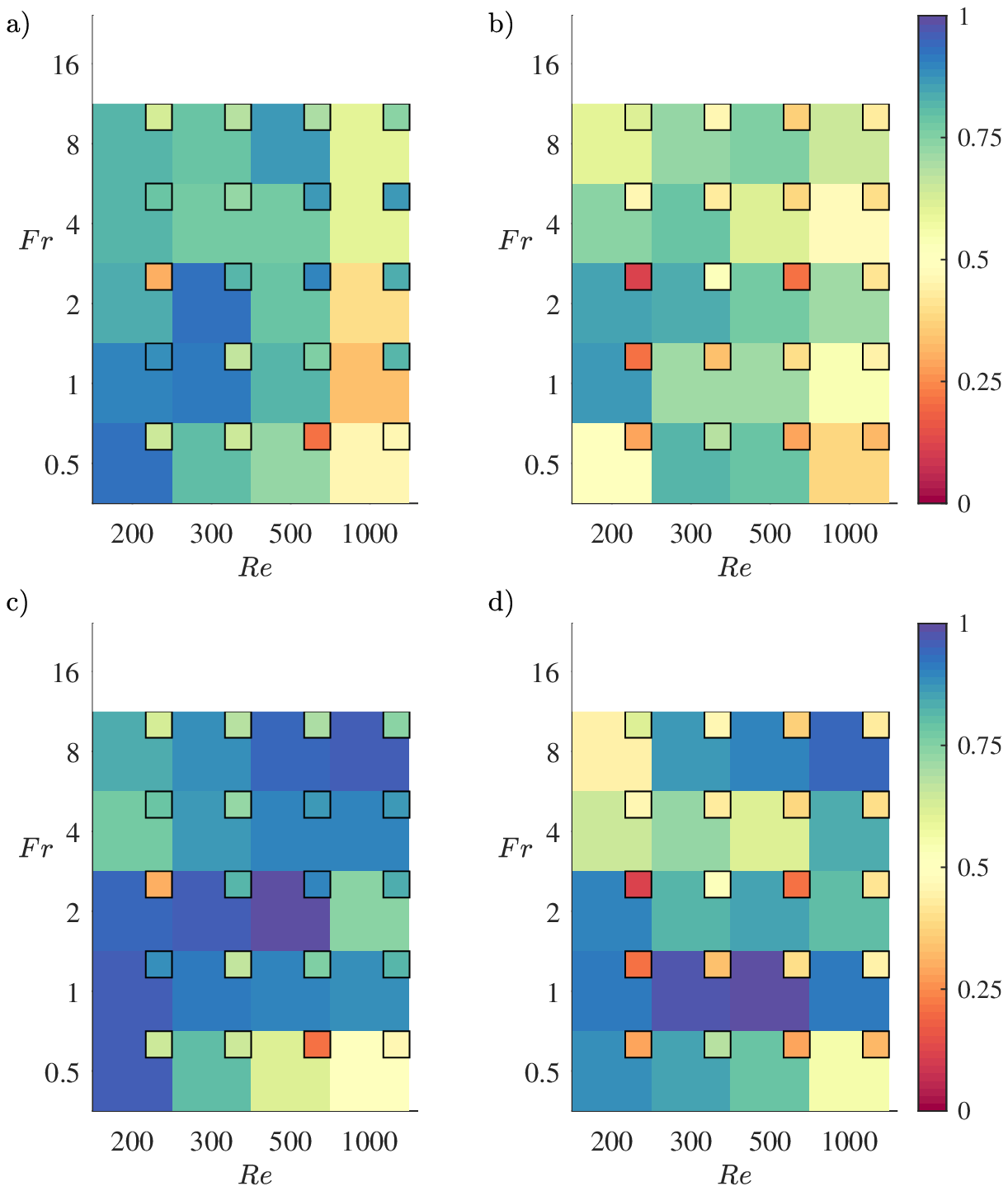}
    \caption{Prediction accuracy \ml{(large boxes, Eq.~\ref{eq:accuracy})} and confidence \ml{(smaller inset boxes, Eq.~\ref{eq:conf})} using experimental measurements of only the vertical component $w$, averaged over 3 different sets of input data. As before, (a) shows results obtained using the full flow fields while (b) shows results obtained using the mean-subtracted flow fields. }
    \label{fig:accuracyPlot_FullDbase_ExpData_2D1C}
\end{figure}

Next, we evaluate whether the method developed here is able to identify parameter regimes using 2D2C PIV snapshots from experiments as input data. As before, we make use of 10 snapshots split into training, testing, and validation data. Compared to regime identification using 2D2C snapshots from DNS, there are three major challenges associated with the use of input data from experiments.  First, the experimental data tend to be noisier.  Second, though the experiments were carried out over the same approximate $(Re,Fr)$ range as the DNS, the $Re$ and $Fr$ values do not match exactly with the DNS cases used to generate the DMD mode database. Third, the spatial extent of the PIV windows in experiment does not coincide with the streamwise extent of the DNS data used for DMD. To remove small-scale noise from the experimental data, we pursue snapshot POD for the experimental data \citep[see e.g.,][]{taira2017modal} and reconstruct the velocity field by retaining the 7 leading POD modes.  In all cases, these leading modes capture more than $99\%$ of the total kinetic energy. We do not take any explicit actions to address the second and third challenges and so the results presented below are indicative of \textit{real-world} reconstruction performance.

Figure~\ref{fig:Re500Fr4MeanFullDbaseExpData} shows reconstruction and regime identification performance using input data from experiments conducted at $(Re,Fr) = (533,3.7)$, similar to the $(Re,Fr) = (500,4)$ case considered in figures~\ref{fig:Re500Fr4MeanFullDbaseNumData} and ~\ref{fig:Re500Fr4MeanSparseDbaseNumData}. The streamwise extent of the wake captured in the experimental data, $x/R \in [26.5,36]$, overlaps partially with the streamwise extent for the DMD modes, $x/R \in [20.25,29.75]$.  Despite the differences in $(Re,Fr)$ values and spatial extent, the method developed here yields reasonable reconstructions for the experimental velocity fields using DMD modes obtained from DNS (see panels (a)-(d) in figure~\ref{fig:Re500Fr4MeanFullDbaseExpData}). The confidence metric, which is a measure of reconstruction accuracy for the test data, is estimated to be approximately $83\%$.  Figure~\ref{fig:Re500Fr4MeanFullDbaseExpData}(f) shows that the regime predictions converge to $(\overline{Re}_{w},\overline{Fr}_{w}) \approx (349, 7.5)$.  At first glance, there is a significant discrepancy between the estimated Froude number and the actual value, $Fr = 3.7$. However, it is important to keep in mind that the DMD mode database spans nearly 2 orders of magnitude in $Fr$. In this context, the algorithm does a reasonable job of localizing parameter values. Moreover, per the classification in figure~\ref{fig:ReFrspace} the estimated $Re-Fr$ combination falls along the same interface between physical regimes as the actual parameter combination.

We next compute the accuracy and confidence for all the available experimental datasets.  Results averaged over three runs are shown in Figure~\ref{fig:accuracyPlot_FullDbase_ExpData}. Note that the PIV FOV for the experimental training data and, hence, the overlap with the DMD modes varies for each case. The wake window is smallest for $(Re,Fr) \approx (1000,0.5)$ with a streamwise extent of $3.125R$ and larger for higher $Fr$ cases.  For cases in which experimental data are available over a larger spatial window, the streamwise extent is capped at $9.5R$, which is the window used for DMD. Reconstruction and regime identification is carried out with the smaller of the two windows (i.e., experimental and DMD) both in streamwise extent as well as in the vertical direction. For some cases there is significant overlap in the streamwise direction (e.g., approximately $8R$ for $(Re,Fr)=(500,2)$) and for others there is little or no overlap (e.g., for $(Re,Fr) = (200,0.5)$). In general, the regime identification accuracy and confidence are lower with the experimental input data compared to training data from DNS. As shown in Table~\ref{tab:summary}, regime identification accuracy is roughly $72\%$ on average with the full flow field from experiments compared to $93\%$ with the DNS snapshots. The lower accuracy and confidence could be attributed to the noise inherent in the experimental data and differences in streamwise extent between the experimental data and the DMD modes. Nevertheless, these results serve as proof-of-concept for regime identification with real-world experimental data.  As before, prediction accuracy is lowest for the high $Re$ cases, and a comparison of the results in figure~\ref{fig:accuracyPlot_FullDbase_ExpData}(a) and (b) shows that predictions improve when the full velocity field from experiments is used as the input rather than just the fluctuations ($72\%$ vs. $68\%$ per Table~\ref{tab:summary}).

It is worth noting here that reconstruction carried out using just the vertical component of velocity ($w$) yields similar results to reconstruction based on both the components of velocity.  This is illustrated by a comparison between figure~\ref{fig:accuracyPlot_FullDbase_ExpData}, which shows results obtained using both components of velocity, and figure~\ref{fig:accuracyPlot_FullDbase_ExpData_2D1C}, which shows results obtained using only the vertical component. As shown in Table~\ref{tab:summary}, overall regime identification accuracy \textit{improves} by roughly $3\%$ with the use of $w$ alone, and there is a nearly $8\%$  improvement in predictions for $Fr$. As was the case with the snapshots from DNS, prediction accuracy is higher for $Fr$ alone compared to the composite metric defined in Eq.~\ref{eq:accuracy}.  Similarly, per Table~\ref{tab:summary}, using a sequence of 30 snapshots from the experiments as the input does not lead to any noticeable improvement in prediction accuracy.

\subsection{Point measurements}\label{sec:TRpoint}

\begin{figure}
    \centering
    \includegraphics[scale=0.9,clip=true]{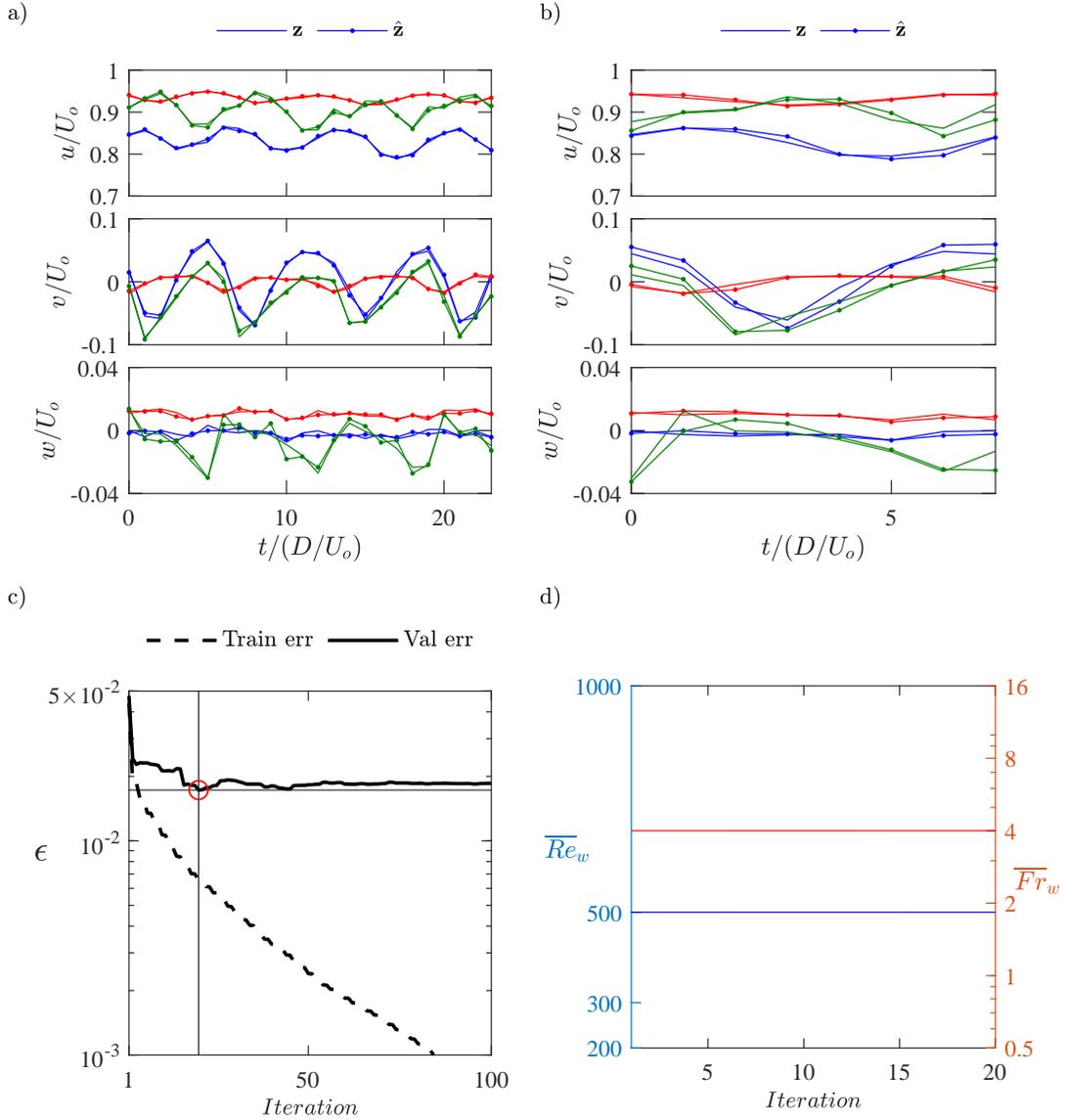}
    \caption{Comparisons between the predictor ($\hat{\mathbf{z}}$) and the DNS ($\mathbf{z}$) over the training (a) and testing (b) datasets. These reconstructions correspond to sensor location 1. The evolution of reconstruction error with the number of iterations for the training and validation sets is shown in (c). The corresponding variation of $\overline{Re}_w$ and $\overline{Fr}_w$ is shown in (d). The input data corresponds to $(Re,Fr)=(500,4)$ and the regime prediction corresponds to $(\overline{Re}_{w},\overline{Fr}_{w}) = (500, 4)$ with a confidence of 97\%.}
    \label{fig:TRdataPlots_fullDbase}
\end{figure}

\begin{figure}
    \centering
    \includegraphics[trim={0cm 7.5cm 0cm 0cm},scale=0.9,clip=true]{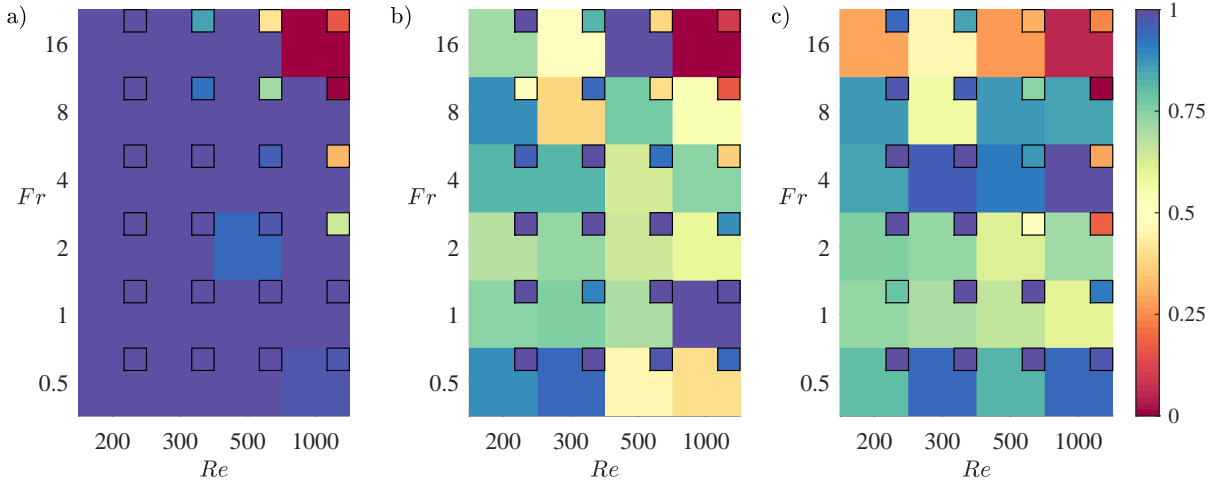}
    \caption{Prediction accuracy \ml{(large boxes, Eq.~\ref{eq:accuracy})} and confidence \ml{(smaller inset boxes, Eq.~\ref{eq:conf})} with point measurements as input data using the full DMD mode database. 
    Subplots (a), (b), and (c) correspond to sensor locations 1, 2, and 3 respectively.
    }
    \label{fig:TRpointMeas_accPlot_FullDbase}
\end{figure}

\begin{figure}
    \centering
    \includegraphics[trim={0cm 7.5cm 0cm 0cm},scale=0.9,clip=true]{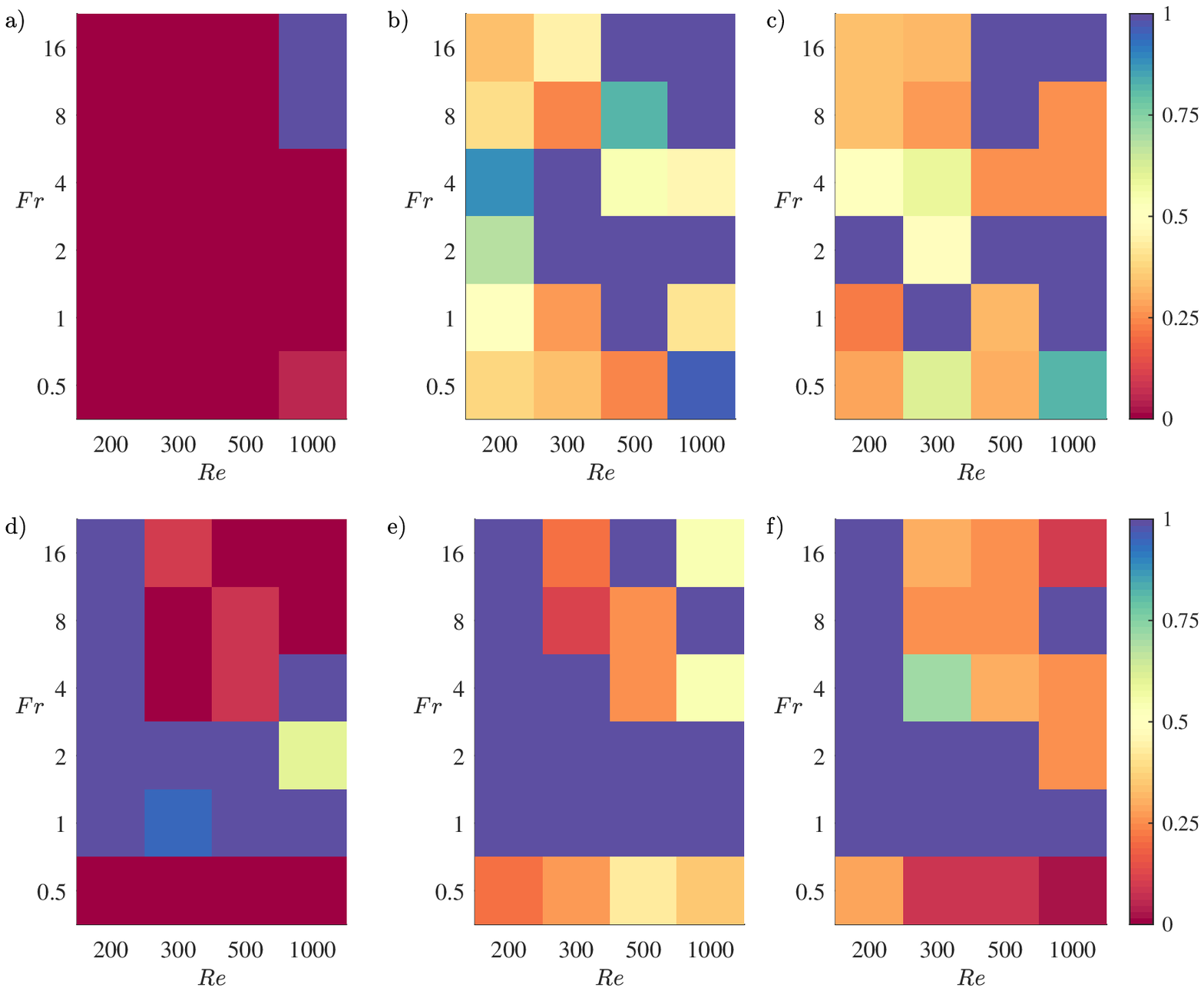}
    \caption{Error in predicted sensor location normalized by $R$.     
    Subplots (a), (b), and (c) correspond to sensor locations 1, 2, and 3 respectively. Note that the error is capped at $1R$ and red shading indicates lower error.}
    \label{fig:TRpointMeas_sensLoc_FullDbase}
\end{figure}

As a final test of the framework developed here, we now attempt regime identification using point measurements (extracted from DNS data).  We present results corresponding to three nominal sensor locations: $(x,y,z)_1=(24,-0.25,-0.25)$, $(x,y,z)_2=(25.875,-0.25,1.25)$, and $(x,y,z)_3=(25.875,1.25,-0.25)$. Recall that these nominal sensor locations represent the central point around which 3 separate point sensors are arranged; the point sensors are respectively placed at a distance of $1.25R$ in the $x$, $y$, and $z$ directions from the nominal sensor location. For example, the coordinates of the three sensors corresponding to sensor location 1 are $(25.25,-0.25,-0.25)$, $(24,-0.25,1)$, and $(24,1,-0.25)$. Sensor location 1 corresponds to measurements made directly downstream of the object.  Sensor locations 2 and 3 are offset in the vertical and spanwise directions, respectively, from the center of the wake. We assume that each point sensor is capable of measuring all three components of velocity. The input data consists of a time series of 40 measurements from each sensor, with 24 measurements dedicated for training, 8 for testing, and 8 for validation. Though we do not pursue an extensive survey of sensor locations, the results obtained in this section provide useful insight into sensor placement for regime identification.

We first start with regime identification based on measurements directly in the wake of the object. Figure~\ref{fig:TRdataPlots_fullDbase} shows comparisons between the point measurements extracted from DNS ($\mathbf{z}$) and the reconstructions ($\hat{\mathbf{z}}$) over the training (a) and testing (b) data for the representative case corresponding to $(Re,Fr)=(500,4)$. The full DMD database is used for reconstruction. Panel (c) shows the evolution in reconstruction error with the number of iterations.  There is a clear minimum in reconstruction error for the validation data after 20 iterations. Panel (d) shows that the predictions converge to $\overline{Re}_{w}=500$ and $\overline{Fr}_{w}=4$ after 20 iterations, yielding a perfect match with the input data.  Recall that the method developed here is also able to estimate the sensor location based on the identified basis functions.  For this case, the identified sensor location matches perfectly with the actual sensor location (see figure~\ref{fig:TRpointMeas_sensLoc_FullDbase}(a)). The close match between the input and reconstructed flow fields for the testing data also yields a high level of confidence ($97\%$) per the metric defined in Eq.~\ref{eq:conf}.

Prediction accuracy and confidence for all 24 $(Re,Fr)$ cases are shown in figure~\ref{fig:TRpointMeas_accPlot_FullDbase} for the three different sensor locations.  As shown in Figure~\ref{fig:TRpointMeas_accPlot_FullDbase}(a), for sensor location 1 the method yields high-accuracy and high-confidence predictions for nearly all cases. Indeed, the average prediction accuracy is over $95\%$ for this scenario (see Table~\ref{tab:summary}). However, the method does not yield accurate predictions for the case with $(Re,Fr)=(1000,16)$.
Figure~\ref{fig:TRpointMeas_sensLoc_FullDbase} shows the normalized error in predicted sensor location. The normalized error is defined as $\max(\Delta/R,1)$, in which $\Delta$ is the distance between the actual and estimated sensor location. Since the wake develops slowly in the streamwise direction, the distance $\Delta$ is computed using the vertical and spanwise coordinates only; the streamwise coordinate is not included. Consistent with the regime predictions, figure~\ref{fig:TRpointMeas_sensLoc_FullDbase}(a) shows that the method is able to identify the correct sensor location in nearly all the cases based on measurements at sensor location 1.  Recall that this location is directly downstream of the body, roughly centered in the wake.

Figures~\ref{fig:TRpointMeas_accPlot_FullDbase}(b,c) show that prediction accuracy and confidence deteriorate for input data from sensor locations 2 and 3.  The average prediction accuracy is roughly $68\%$ for sensor location 2 and $71\%$ for sensor location 3 (c.f., $95\%$ for location 1; see Table~\ref{tab:summary}). Figures~\ref{fig:TRpointMeas_sensLoc_FullDbase}(b,c) show a similar deterioration in sensor location estimates.  This deterioration of performance is perhaps not surprising given that sensor locations 2 and 3 are offset from the center of the wake. Interestingly, the use of mean-subtracted flow fields alleviates some of the performance degradation in regime identification from measurements at sensor location 2. Table~\ref{tab:summary} shows that the average prediction accuracy \textit{improves} from $68\%$ to $77\%$ if only the velocity fluctuations from sensor location 2 are used for regime identification. The improvement of performance with the use of mean-subtracted flow fields is also evident for sensor location 3; average prediction accuracy \textit{improves} from roughly $71\%$ for the full flow field to $77\%$ for fluctuations alone.  In general, the results presented in Table~\ref{tab:summary} show a smaller variation in prediction accuracy across the different sensor locations when mean-subtracted flow fields are used as input data. This suggests that the mean velocity field is useful for regime identification from point measurements, but primarily for measurements made close to the wake centerline.  For measurements made away from the center, the mean component may have a confounding influence.

For completeness, we note that we also pursued regime identification from point measurements using the sub-sampled DMD mode database.  Similar to the trends observed for 2D2C snapshots, the use of the smaller database led to a reduction in overall prediction accuracy. Further, prediction accuracy was generally better for just $Fr$ compared to the composite $Re-Fr$ metric defined in Eq.~\ref{eq:accuracy}.  These results are summarized in Table~\ref{tab:summary}. 

\section{Conclusions}\label{sec:conclusions}

In this paper, we developed a data-driven framework for identifying the dynamic regime (i.e., $Re-Fr$ values) for stratified wakes from limited velocity snapshots or point measurements.  This approach makes use of a library of DMD modes available from prior DNS.  The FROLS algorithm is used to identify DMD modes that best represent the input data, and $Re-Fr$ are then estimated via a projection-weighted average of parameter values for the selected DMD modes. We demonstrated the utility of this framework using input data from both numerical simulations and laboratory experiments. Regime identification accuracy was very high ($>90\%$; see Section~\ref{sec:2D2C-DNS}) when velocity snapshots from numerical simulations were used as input data.  Moreover, prediction accuracy deteriorated only slightly when the library (or database) of DMD modes was sub-sampled. This confirms that the method is capable of interpolating between parameter values when the specific $Re-Fr$ combination is not included in the library.  The average regime identification accuracy from velocity snapshots obtained in laboratory experiments was lower ($>70\%$; see Section~\ref{sec:2D2C-EXP}). This deterioration in performance was expected given that the laboratory measurements were inherently noisier and did not have the same spatial extent as the DMD modes obtained from numerical simulations. Given these challenges, the reasonable prediction accuracy suggests that this technique is robust and could be used for real-world regime identification. The results presented in Section~\ref{sec:TRpoint} show that this framework is capable of regime identification from point measurements. When using point measurements, the method also yields reasonable estimates for sensor location. 

\ml{We recognize that the parameter combinations considered here correspond to low Reynolds number flows that are relatively narrow-banded. Regime identification is expected to become more challenging for higher $Re$ flows that exhibit a wider range of spatiotemporal scales. In other words, the limited datasets typically obtained from laboratory and field measurements may not contain enough information to allow for meaningful delineation between regimes at higher $Re$. Even in the present study, a deterioration in performance is evident at $Re=1000$ for inputs from both numerical simulations and experimental measurements.  Yet another challenge for high $Re$ applications is the limited availability of high-fidelity simulation and experimental data. This would restrict the library of candidate basis functions that can be used for regression. However, these challenges associated with data scarcity can be overcome with the development of better measurement systems, the use of emerging techniques for optimal sensor placement and data fusion \citep{manohar2018data,wang2021model}, and continued advances in our ability to simulate higher Reynolds number flows.}

In many ways, this study is complementary to the neural network based classifiers \citep[e.g.,][]{colvert2018classifying,alsalman2018training} and expert-defined decision trees \citep[e.g.,][]{ohh2020automated,ohh2021wake} developed in recent work on wake flows. However, the approach developed here has some advantages. First, the library-based framework developed here is flexible and can be used with different forms of input data (e.g., snapshots or point measurements with varying spatiotemporal fidelity). Feed-forward neural networks must be developed and trained independently for different input data and similarly decision trees must be modified to account for differing data types. Second, in addition to classifying the data into different regimes, the framework developed here can also reconstruct the velocity field. The confidence in the constructed model can therefore be evaluated based on how well \ml{the testing data are predicted.}

Note that the framework developed here is not limited to the use of DMD modes. The FROLS algorithm simply identifies the best basis functions that represent the data.  These basis functions can be Fourier modes, DMD modes, POD modes, or indeed, any set of candidate basis functions that is sufficiently rich to characterize the input data. Previous work shows that, in addition to DMD, Spectral POD (SPOD; \cite{towne2018spectral}) can also be used to characterize different wakes \citep{nidhan2019dynamic, ohh2021wake}. Additional work is needed to see if the use of SPOD modes improves regime identification and reconstruction accuracy for stratified wakes. Of course, any such evaluation must also account for the increased data requirements for SPOD, which relies on ensemble averaging. Future work also includes a systematic evaluation of reconstruction accuracy as a function of spatial and/or temporal resolution of the input data as well as the duration of the available time series. Nevertheless, this manuscript shows that the library-based sparse regression formulation developed here holds promise for the development of automated data-driven fluid pattern classifiers.

\begin{acknowledgments}
Support for C.Y. Ohh and G.R. Spedding from ONR grant\# N00014-20-1-2584 under Dr Peter Chang
is most gratefully acknowledged.
\end{acknowledgments}

\bibliographystyle{unsrtnat}
\bibliography{Bibliography}
\end{document}